\documentclass[11pt,a4paper]{article}
\usepackage[T1]{fontenc}
\usepackage[utf8]{inputenc}
\usepackage[english]{babel}
\usepackage[margin=24mm]{geometry}
\usepackage{newpxtext,newpxmath}
\usepackage{microtype}
\usepackage{graphicx}
\usepackage{amsmath}
\usepackage{multirow,adjustbox}
\usepackage[font=small,labelfont=bf,labelsep=period]{caption}
\usepackage[authoryear,round]{natbib}
\usepackage{xcolor}
\definecolor{accent}{HTML}{264B60}
\usepackage{titlesec}
\titleformat{\section}{\large\sffamily\bfseries\color{accent}}{\thesection}{0.7em}{}
\titleformat{\subsection}{\normalsize\sffamily\bfseries}{\thesubsection}{0.7em}{}
\usepackage{hyperref}
\hypersetup{colorlinks=true,linkcolor=accent,citecolor=accent,urlcolor=accent,
 pdftitle={Other detection methods},pdfauthor={Jorge Lillo-Box and Olga Balsalobre-Ruza}}
\begin{document}
\begin{flushleft}
{\LARGE\sffamily\bfseries Exoplanet detection through \\ photometric orbital phase modulations\par}
\vspace{1em}
{\large Jorge Lillo-Box \& Olga Balsalobre-Ruza\par}
\vspace{0.5em}
{\small Centro de Astrobiolog\'ia (CAB), CSIC-INTA,\\
Camino Bajo del Castillo s/n, 28692,\\
Villanueva de la Ca\~nada (Madrid), Spain\par}
\end{flushleft}
\vspace{0.3em}
\noindent\textcolor{accent}{\rule{\linewidth}{0.6pt}}
\begin{abstract}
Since the detection of the first extrasolar planets thirty-five years ago, the dawn in detected signals has been driven primarily by radial velocity and transit techniques. However, other methodologies have gained significant interest due to their ability to explore specific niches that are challenging to access with traditional approaches. In this article, we focus on the photometric orbital phase modulations induced by planets, describe the methodologies and equations and highlight their developments and prospects for the future.
\end{abstract}

\section{Introduction}

Periodic variations in stellar light curves can reveal the presence of close-in planets, even if they do not transit their host star.
Since the integrated photometry captures the combined light from both the star and the planet, oscillations can arise from the starlight reflected by the planetary atmosphere throughout the orbit (e.g., \citealt{charbonneau99, collier-cameron99}).
Additionally, the star-planet gravitational interactions can modulate the observed stellar flux as a result of ellipsoidal deformations on the external layers of the stellar atmosphere (e.g., \citealt{borucki09, welsh10}), and the photometric counterpart of the radial velocity effect (the so-called Doppler beaming or boosting, e.g., \citealt{barbier21}).
The REB methodology consists in detecting either of these three effects, with its name serving as an acronym for Reflection, Ellipsoidal, and Beaming.

The study of photometric variability in multiple-star systems over the past century provided the mathematical framework to understand the different effects.
The REB effects were particularly well understood in the context of close binaries (e.g., \citealt{morris85, shakura87, wilson90}).
But they have also served to identify other types of close stellar companions, for instance black holes (e.g., \citealt{greene01}), central stars of planetary nebulae (e.g., \citealt{bond00, miszalski09, aller20}), and even sub-stellar objects (e.g., \citealt{pfahl08, lillo-box21b}).

The advent of high-precision, space-based photometric missions nearly two decades ago revolutionized the exoplanetary exploration.
The MOST \citep{gordon03} and CoRoT \citep{auvergne09} missions, the subsequent NASA planet hunter \textit{Kepler} \citep{borucki10} and its extended \textit{K2} mission \citep{howell14}, as well as its successor \textit{TESS} \citep{ricker14},
have surveyed the sky, searching for the dimming events imprinted in the light curves of stellar sources. Soon, the PLATO mission \citep{rauer14} will also enter the scene.
To 2025, more than 6000 signals consistent with a planetary origin have been detected around stars, with approximately three-fourths of them attributed to the transit method.
However, transits are not the only profit we can make of long-term, high-cadence, photometric light curves. Motivated by the exceptional quality of the Kepler data and drawing on the techniques used in binary star systems, \cite{faigler11} proposed the REB variations as an alternative method for detecting new planets.

Since the ellipsoidal variations and Doppler beaming effect are gravitationally driven, they depend on the planetary mass.
Consequently, this technique can be used not only to detect planets, but also, to potentially confirm their planetary nature if the measured mass is below 13\,M$_{\rm Jup}$ \citep{spiegel11}.
Additionally, it enables the full characterization when both radius and mass can be estimated (e.g., \citealt{mazeh12, esteves13, lillo-box14, millholland17}). Importantly, taking ellipsoidal and beaming effects into account avoids biasing the results when studying reflected and thermal phase variations.

The REB modulations have already been used to detect new non-transiting planets, as it is the case of
Kepler-76\,b \citep{faigler13}, KIC~5479689\,b, KIC~8121913\,b, KIC~10068024\,b \citep{lillo-box21}, and other several tens of \textit{Kepler} candidates \citep{millholland17}. They have been also applied to characterize almost thirty known close-in transiting planets, such as Kepler-5\,b, Kepler-6\,b, Kepler-8\,b, Kepler-13\,b, KOI-64\,b, TrES-2\,b, HAT-P-7\,b \citep{esteves13}, Kepler-91\,b \citep{lillo-box14, placek15}, and Kepler-13\,b \citep{shporer14, placek14}, among others\footnote{A more complete list can be found in the \href{https://exoplanetarchive.ipac.caltech.edu/cgi-bin/TblView/nph-tblView?app=ExoTbls&config=PS}{NASA Exoplanet Archive} table using the \textit{detected by Orbital Brightness Modulations} filter.}. Those systematic searches have established several interesting findings, including generally low (but occasionally quite large) hot Jupiter albedos \citep{demory11} and repeated detections of deviations in the maximum of the phase curve from the sub-stellar point (e.g., \citealt{demory13, shporer15, millholland17}).

In this section, we will describe the three most relevant sources of light curve modulation above mentioned in the context of a planet-star system. We will discuss their dependencies, and their implications in the exoplanetary field. We focus our description in the optical regime but acknowledge that additional effects may be more relevant in other regimes (e.g. tidal distortion of the exoplanet shape in the infrared - \citealt{barros22}-, or planetary thermal emission against reflected light - \citealt{cowan12}).

\section{Method description}
\label{sec:method}

The total contribution to the orbital phase modulations is the combination of the planetary component, including both reflected and thermal emissions, $(\Delta F/F)_{\rm ref}$ and $(\Delta F/F)_{\rm th}$, the ellipsoidal distortion, $(\Delta F/F)_{\rm ell}$, and the Doppler beaming, $(\Delta F/F)_{\rm beam}$. Therefore

\begin{equation}
\label{eq:general}
	\left(\frac{\Delta F}{F} \right) = \left( \frac{\Delta F}{F} \right)_{\rm ref} +
									  \left( \frac{\Delta F}{F} \right)_{\rm th} +
									  \left( \frac{\Delta F}{F} \right)_{\rm ell} +
									  \left( \frac{\Delta F}{F} \right)_{\rm beam}.
\end{equation}

In the following subsections we describe the mathematical form of each term and discuss the potential degeneracies between them.

\subsection{Planetary light reflection and thermal emission}

\label{sec:reflection}

The flux received from the planet is compound of starlight reflected by its surface or atmosphere, and thermal radiation emitted directly by the planet.
The former effect results from the stellar radiation striking the day-side of the planet, with a portion of it being reflected depending on the planetary albedo.
On the other hand, thermal emission is caused by the planetary temperature, which is driven by internal heating and stellar energy absorption.
These two emissions contribute to the overall flux that we receive in our detectors, with the thermal emission typically becoming significant only at longer wavelengths and short periods.

It is intuitive that the planetary contribution to the light curve modulations depends on the the fraction of the planet's projected disk illuminated by the star.
As seen from  the observer's direction, this fraction is the so-called \textbf{phase function}, $\Phi(z)$, which typically adopts two definitions.
The most simple one is assuming a specular reflector for the day-side of the planet, in which the received flux is proportional to the projected illuminated area of the planet on the plane of the sky.
In this case, it is called the \textbf{geometrical phase function}, $\Phi_{\rm geo}(z)$.
The alternative formulation assumes that the surface of the planet behaves as an ideal isotropic reflector, with the hemisphere facing the observer scattering light equally in all directions (known as a Lambert sphere, \citealt{lambert1760,russell1916}).
This assumption is referred to as the \textbf{Lambertian reflection phase function}, $\Phi_{\rm Lam}(z)$, and it is the most commonly used in the analysis of light curve modulations of extrasolar planets.

In both cases, the phase function depends on the so-called \textbf{\emph{z}} \textbf{angle} (with $0 < z \leq 2\pi $), which is the angle between the star and the observer as seen from the planet (i.e., the star-planet-observer angle)\footnote{Note that with this definition, $z=0$ corresponds to the inferior conjunction of the planet, (i.e., when the planet is perfectly aligned with the star and the observer passing between both), while $z=\pi$ corresponds to the superior conjunction (i.e., when the planet aligns with the star passing behind it).}. The \emph{z} angle is related to the \textbf{geometric phase angle} ($\theta$) and the orbital inclination ($i$) such that

\begin{equation}
\label{eq:z_theta}
\cos{z(t)} = -\sin{i}\cos{\theta(t)}.
\end{equation}

Note that the geometric phase angle is only equal (with a 2$\pi$ factor) to the temporal phase angle, $\phi(t)$, for circular orbits (zero eccentricity, $e = 0$), being

\begin{equation}
\label{eq:temporal_ph_angle}
\theta(t) = 2 \pi \phi(t) = 2\pi(t-T_0)/P,
\end{equation}
where $T_0$ is the time of inferior conjuction, and $P$ the orbital period of the planet. Otherwise ($e \neq 0$), the relation is provided by solving the Kepler's equation.

Therefore, the geometrical phase function case is described as the cosine of the geometrical angle between the line of sight and the position of the planet on its orbit, being

\begin{equation}
\label{eq:phasefunc_geo}
\boldsymbol{\Phi_{\rm geo}(z)} = \cos{z} = -\sin{i}\cos{\theta}.
\end{equation}

While the Lambertian reflection phase function is

 \begin{equation}
\label{eq:phasefunc_Lam}
\boldsymbol{\Phi_{\rm Lam} (z)} = 2 \frac{ (\sin{|z|}+(\pi-|z|)\cos{|z|})}{\pi}.
\end{equation}

As demonstrated by \cite{faigler14}, expanding $\Phi_{\rm Lam}(z)$ in a Fourier series and using the relation between $z$ and $\theta$, we find

\begin{equation}
\label{eq:phasefunc_LamFourier}
\Phi_{\rm Lam} (\theta) = -\sin{i}~\boldsymbol{\cos{(\theta)}}  + 0.18\sin^2i ~\boldsymbol{\cos{(2\theta)}} + O(\sin^3{i}).
\end{equation}

Interestingly, the first harmonic is the geometric phase function defined in Eq.~\ref{eq:phasefunc_geo} and the second harmonic has a $\cos{(2\theta)}$ dependency.
Consequently, the second harmonic can contribute by increasing up to 18\% (or smaller for inclined orbits) the geometric term (assuming a uniform albedo, which might not be the case for short period planets due to non-uniform cloud coverage, e.g., \citealt{demory13}). The $\cos{(2\theta)}$ dependency might be a source of degeneracy when trying to derive the mass of the perturber (see Sect.~\ref{sec:REBtesting}).

Once the phase function is defined, we can analyze the direct contribution of the planet to the modulation of the light curve:

\begin{enumerate}
\item \textbf{Reflected light.}
The ratio between the incident stellar flux and the reflected light by the planetary day side
is described by the geometric albedo of the planet, $A_g$\footnote{An albedo of zero corresponds to a black body that absorbs all incident light, whereas an albedo of one represents a perfect mirror that reflects it entirely.}.
If we call $F_0$ to the incident flux at the planetary distance ($r = \frac{a (1-e^2)}{1+e\cos{\nu}}$), the relative reflected light would be $(F_0 A_g \pi R_p^2) /(4\pi r^2) \Phi(z)$, where $R_p$ is the planetary radius.
Hence, the contribution to the light curve modulation of the reflected light is

\begin{equation}
\label{eq:reflection}
\left( \frac{\Delta F}{F} \right)_{\rm ref} = A_g \left(\frac{R_p}{r}\right)^2 \Phi(z)  ~ \equiv ~ - A_{\rm ref} \Phi(z).
\end{equation}

The albedo is strongly dependent on the wavelength, hence it is often expressed as $A_g(\lambda)$. It has been also demonstrated to depend on the host metallicity, and star-planet separation \citep[see, for example, ][]{sudarsky05, cahoy10}.
Several authors have worked on deriving analytic functions to describe the geometric albedo, although this is a complex task without a general form.
For instance, \cite{madhusudhan12} and \cite{heng21} provided a comprehensive (but complex) scheme to compute the geometric albedo and phase function accounting for various types of scattering in planetary atmospheres, including isotropic, asymmetric, Lambertian, and Rayleigh scattering.
A simple formulation for giant extrasolar planets was provided by \cite{kane10} with the albedo being an hyperbolic tangential function depending only on the planet-star separation, based on the theoretical models of \cite{sudarsky05}

\begin{equation}
\label{eq:ag}
 A_g=\frac{e^{r-1}-e^{-(r-1)}}{5(e^{r-1}+e^{-(r-1)})} + \frac{3}{10},
\end{equation}

\noindent where the planet-star separation is in astronomical units (AU).
It has been demonstrated to work for several Solar System planets.
However, it provides only an upper limit for the albedo of short-period hot-Jupiter planets, as its lower bound is $A_g = 0.147$ but many planets have been found to exhibit lower values.
In practice, typically the albedo is left as a free parameter when modeling this effect, where Eq.~\ref{eq:ag} can be used as an a priori estimation. Indeed, in case that the planet transits its host star and under the assumption of a particular phase function (Lambertian of geometric) and uniform albedo, $A_g$ is the only unknown parameter from Eq.~\ref{eq:reflection}.
If the secondary eclipse is also detected with a depth $\Delta F_{\rm ecl}$, the geometric albedo can be directly determined as the phase function at superior conjunction is $\Phi(z=\pi)=1$, hence Eq.~\ref{eq:reflection} becoming $\Delta F_{\rm ecl} = A_g(R_p/r)^2$ (neglecting the thermal contribution of the planet).

\item \textbf{Thermal emission.}
The contribution to the total flux of the system from both the day- and night-sides was described by \cite{cowan11} and can be formulated as

\begin{equation}
\label{eq:thermal_day}
\left( \frac{\Delta F}{F} \right)_{\rm th,day}  = \Phi(z) \left(\frac{R_p}{R_{\star}}\right)^2 \frac{e^{hc/\lambda k T_{\rm eff} }-1}{e^{hc/\lambda k T_{\rm day} }-1},
\end{equation}

and

\begin{equation}
\label{eq:thermal_night}
\left( \frac{\Delta F}{F} \right)_{\rm th,night}  = \left[1-\Phi(z) \right]~ \left(\frac{R_p}{R_{\star}}\right)^2 \frac{e^{hc/\lambda k T_{\rm eff} }-1}{e^{hc/\lambda k T_{\rm night} }-1},   \\
\end{equation}

where $R_\star$ is the stellar radius, $h$ is Planck’s constant, $c$ is the light speed, $\lambda$ is the effective wavelength of the bandpass (e.g., $5750$ \AA\ in the case of \emph{Kepler}, see \emph{Kepler} Handbook\footnote{\url{https://archive.stsci.edu/kepler/manuals/KSCI-19033-001.pdf}}), $k$ is the Boltzmann constant, and $T_{\rm eff}$ is the stellar effective temperature. The day and night temperatures are defined by

\begin{equation}
\label{eq:day_temp}
T_{\rm day} = T_{\rm eff} \left(\frac{r}{R_{\star}}\right)^{-1/2} (1-\alpha_{\rm bol})^{1/4}\left( \frac{2}{3}- \frac{5}{12}\epsilon \right)^{1/4},
\end{equation}

and

\begin{equation}
\label{eq:night_temp}
T_{\rm night} = T_{\rm eff} \left(\frac{r}{R_{\star}}\right)^{-1/2} (1-\alpha_{\rm bol})^{1/4}\left(\frac{\epsilon}{4} \right)^{1/4}.
\end{equation}

In these equations, $\alpha_{\rm bol}$ is the bolometric albedo of the planet, and $\epsilon$ is the energy circulation across the exoplanet atmosphere ranging from 0 to 1, where $\epsilon=0$ corresponds to the maximum thermal emission.

It is important to note that Eqs.~\ref{eq:thermal_day} and \ref{eq:thermal_night} are first-order approximations that do not account for other effects like phase offsets or high-order harmonics \citep[e.g.,][]{cowan13}. As stated by \cite{mislis12}, the thermal emission contribution at optical wavelengths for hot Jupiter planets is significantly smaller than the reflection component of the modulation, accounting for less than 1\% when orbital periods exceed 2 days around stars with $T_{\rm eff}<6500$~K (see Figs. 3 and 4 in the aforementioned paper).

\end{enumerate}

With all these definitions, the relative planetary contribution to the total light curve is given by

\begin{equation}
\label{eq:planet_total}
\left( \frac{\Delta F}{F} \right)_{\rm planet} = \left( \frac{\Delta F}{F} \right)_{\rm ref} +
												 \left( \frac{\Delta F}{F} \right)_{\rm th,day} +
												 \left( \frac{\Delta F}{F} \right)_{\rm th,night}.
\end{equation}

The overall dependence of these effects on the orbital position of the planet is as $\Phi(\theta) \propto -\cos{\theta}$, except for the less significant effect of the thermal emission from the night side of the planet which varies as  $+\cos{\theta}$.
Hence, the planetary contribution is maximized at opposition ($\theta =\pi$) where the day side is fully visible, and minimized at conjunction ($\theta=0$) when the night side is observed.

\subsection{Ellipsoidal variations}

\label{sec:ellipsoidal}

Ellipsoidal variations arise from tidal effects caused by the gravitational interaction between a companion and its host star.
The companion acts as a gravitational perturber altering the spherical shape of the stellar surface.
In the simplest model, the stellar surface adopts a prolate ellipsoidal shape that co-moves with the companion, with its major axis aligned with the direction connecting the two objects.
As the planet orbits, the varying projected area of the star in the plane of the sky leads to changes in the amount of light reaching the observer, resulting in detectable variability in the measured stellar flux.

These variations were theoretically studied by \cite{pfahl08}, who analytically determined the flux changes caused by stellar oscillations induced by sub-stellar companions.
The derived equations assume Keplerian orbits, slow stellar rotation such that only radial surface perturbations are significant, spin-orbit alignment (i.e., the orbital plane is perpendicular to the stellar rotation axis), and neglect the contribution from Doppler shifts due to wave motions on the stellar surface.
Under these assumptions, the relative variability in the bolometric stellar flux received from the star is expressed as:

\begin{equation}
\label{eq:pfahl}
\left(\frac{\Delta F}{F} \right)_{\rm ell} = \frac{M_p}{M_{\star}} \left(\frac{R_{\star}}{a} \right)^3 \, \sum_{l=2}^{+\infty} \left( \frac{R_{\star}}{a} \right)^{l-2}  \left( \frac{a}{r}  \right)^{l+1} f_l P_l(\cos{\psi_o}),
\end{equation}

\noindent where $M_p / M_\star$ is the planet-star mass ratio, $a$ is the orbit semi-major axis, and $r$ is the companion-star distance, with $e$ the eccentricity and $\nu$ the true anomaly. For Eq.~\ref{eq:pfahl} we also need to define $f_l = (2-\Lambda_l)b_l - c_l$,
with $\Lambda_l=l+2$ being a good approximation for radiative stars, and $b_l$ and $c_l$ being functions of the limb darkening coefficients \citep[see Eq. (9) in ][]{pfahl08}. For $l=2$ and $l=3$, these parameters can be written as

\begin{align}
\label{eq:LDcoeff}
b_2 & = \frac{1+\gamma}{20(3-\gamma)}
&  b_3   & = \frac{\gamma}{4(3-\gamma)}  \\
c_2 & = \frac{3(1+3\gamma)}{10(3-\gamma)}       &   c_3  & = \frac{3\gamma}{3-\gamma},
\end{align}

where $\gamma$ is the linear limb darkening coefficient, which is tabulated for different bands by \cite{claret11}.
In Eq.~\ref{eq:pfahl}, $P_l(\cos{\psi_o})$ are the well-known Legendre polynomials depending on $\psi_o$, which is a function of the orbital inclination and the geometric phase angle. For the second ($l=2$) and third ($l=3$) harmonics they can be written as

\begin{equation}
\begin{split}
\label{eq:legendre}
P_2 & = \frac{1}{4} \left[ -(3\cos^2{i}-1) + 3\sin^2{i}\  \boldsymbol{\cos{\left( 2\theta \right)}} \right]  \\
P_3 & = \frac{1}{8}\sin{i} \left[ -3(5\cos^2{i}-1) \boldsymbol{\cos{\theta}} + 5\sin^2{i}\ \boldsymbol{\cos{\left( 3\theta\right)}} \right].
\end{split}
\end{equation}

In Fig.~\ref{fig:L2L3}, we show the contributions of the second and third harmonics to illustrate these equations. In this example, we consider several orbital configurations $\{e,\omega\}$, with $\omega$ being the argument of periastron. We assume a Jupiter-mass planet around a solar-mass star with $a/R_{\star}=5$, $\gamma=3/5$, and an edge-on orbit (i.e., $i=\pi/2$).  Fig.~\ref{fig:ellip_harmonics} shows the equipotential surface filled by the external layers of the star as the companion moves along the orbit. These analyses of the harmonics (presented in Eq.~\ref{eq:pfahl}) reveal the following:

\begin{figure}[htbp]
\centering
\includegraphics[width=1\textwidth]{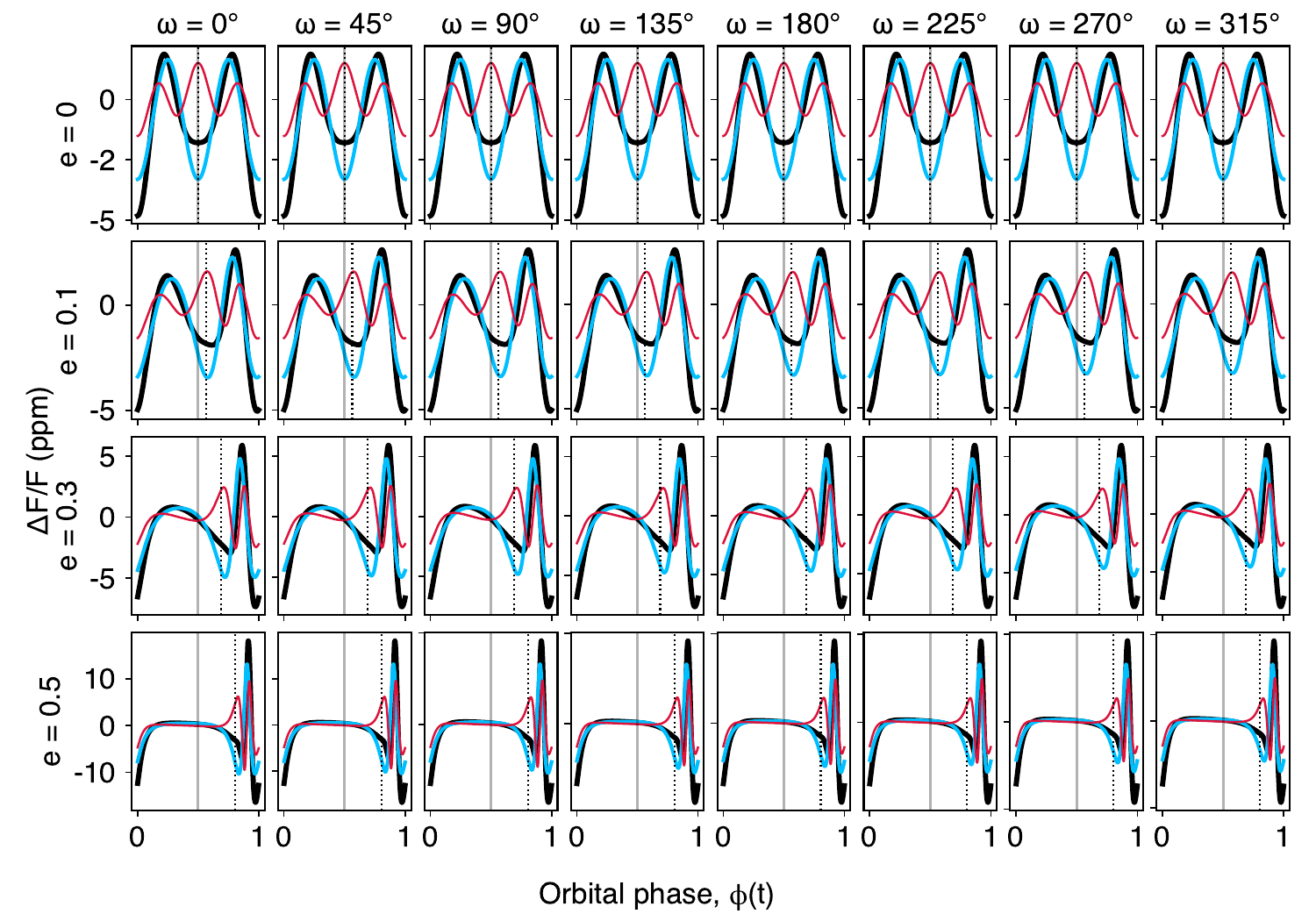}
\caption{Ellipsoidal variations for different orbital configurations of a Jupiter-mass planet around a solar-mass star, and assuming $a/R_{\star}=5$, $\gamma=3/5$, $i=\pi/2$.
The colored lines represent the total contribution (black), and the second (blue) and third (red) harmonics. The vertical lines mark the location of the planet in opposition (solid) and the mid-period (dotted).}
\label{fig:L2L3}
\end{figure}

\begin{figure}[htbp]
\centering
\includegraphics[width=1\textwidth]{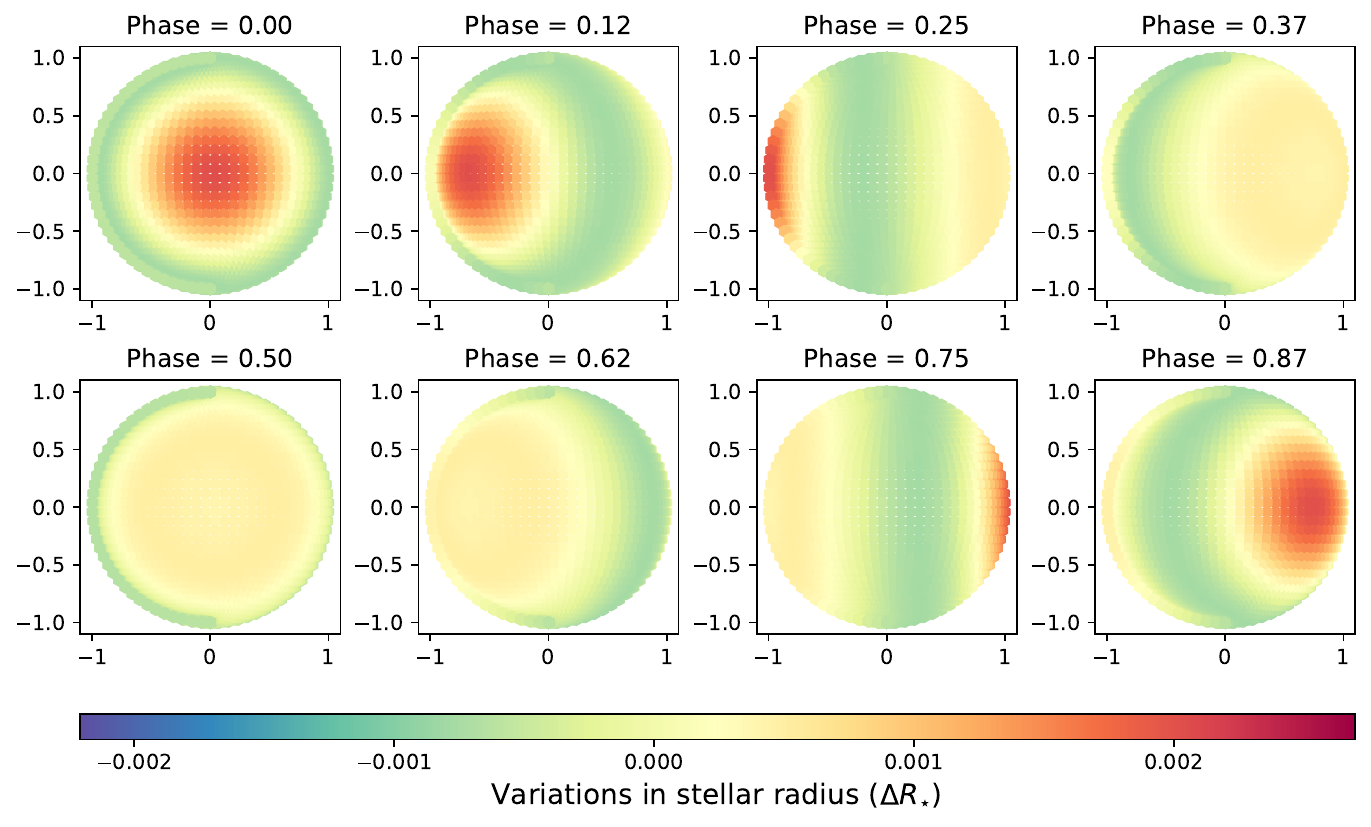}
\caption[Ellipsoidal shape of the stellar outer layers at different positions of the companion]{Ellipsoidal shape (including second and third harmonics) of outer layers of the star at different positions of the companion along its eccentric orbit ($e=0.3$, $\omega=90^{\circ}$, $i=\pi/2$). The color-code represents the radius of the star with respect to its nominal value at the different latitudes and longitudes.}
\label{fig:ellip_harmonics}
\end{figure}

\begin{itemize}
\item \textbf{Second harmonic ($l=2$)}. It dominates the flux variation since $R_{\star}/a<1$ and the exponent increases with $l-2$.
According to the corresponding Legendre polynomial, this harmonic behaves as $\cos{2\theta}$.
In the case of a circular orbit, this implies a modulation of the stellar flux with a periodicity of $P_{\rm orb}/2$, with maxima occurring at $\phi(t)=0.25$ ($\theta=\pi/2$) and $\phi(t)=0.75$ ($\theta=3\pi/2$), as shown in Fig.~\ref{fig:L2L3}.
For non-circular orbits, the modulation becomes asymmetric in the phase-folded light curve.

\item \textbf{Third harmonic ($l=3$)}.
It includes dependencies with the geometrical phase of the companion as $\cos{\theta}$ and $\cos{3\theta}$.
In the particular case of circular orbits, the third harmonic increases the flux at $\phi(t)=0.5$ ($\theta=\pi$), while decreasing at $\theta(t)=\phi(t)=0$ (i.e., during the primary transit).
Additionally, the two maxima are slightly shifted in phase, moving towards $\phi(t)<0.25$ and $\phi(t)>0.75$.
As noted by \cite{pfahl08}, the third harmonic also plays an important role in non-circular orbits.
It becomes more significant as the planet gets closer to the star (i.e., decreasing $a/R_{\star}$).
When including this harmonic, asymmetries from the prolate ellipsoid appear (see Fig.~\ref{fig:ellip_harmonics}).
They are clearly visible at $\theta(t)=90^{\circ}$ and $\theta(t)=270^{\circ}$, where the region of the star facing the companion is more inflated than the opposite side, resembling a tear drop.
\end{itemize}

Interestingly, as far as the orbit is eccentric, the ellipsoidal variations can be detected at any orbital inclination.
Hence, this detection technique could be useful for detecting non-transiting companions in near face-on orbits ($i\sim 0^{\circ}$).

Based on this general formulation by \cite{pfahl08}, several models have been proposed to describe the observed flux variability from candidate host stars with different properties.
Table~\ref{tab:ellipsoidal} summarizes the most commonly used formalisms, highlighting the differences in both amplitude and phase dependencies according to each author.
Most of them use the same amplitude, which only holds for circular orbits.
This expression is derived by substituting in Eq.~\ref{eq:pfahl} all the definitions described above. Hence, we get the model for the ellipsoidal variations as

\begin{equation}
\label{eq:ellip_all}
\begin{split}
\left(\frac{\Delta F}{F} \right)_{\rm ell} =  \frac{M_p}{M_{\star}} ~ \left( \frac{R_{\star}}{r} \right)^3 ~   \overbrace{ \left(  \frac{1}{4} f_2\left[ -(3\cos^2{i}-1) + 3\sin^2{i}\  \boldsymbol{ cos{\left( 2\theta \right)}} \right] \right.}^{l=2} + \\
   + \underbrace{ \left. \frac{1}{8}\left( \frac{R_{\star}}{r} \right) f_3 \sin{i} \left[ -3(5\cos^2{i}-1) \boldsymbol{\cos{\theta}} + 5\sin^2{i}\ \boldsymbol{\cos{\left( 3\theta \right)}} \right]  \right) }_{l=3}.
\end{split}
\end{equation}

Neglecting the third harmonic order (which can be highly inaccurate for specific cases) and expanding the remaining equation, we find

\begin{equation}
\label{eq:ellip_2nd}
\begin{split}
\left( \frac{\Delta F}{F} \right)_{\rm ell} \approx \left[ \frac{3f_2}{4}  ~ \frac{M_p}{M_{\star}} ~ \left( \frac{R_{\star}}{r} \right)^3 ~  \sin^2{i}\  \boldsymbol{ \cos{\left(2\theta \right)}} \right] - \\
                            - \left[ \frac{M_p}{M_{\star}} ~ \left( \frac{R_{\star}}{r} \right)^3 ~ \frac{f_2}{8}  (3\cos^2{i}-1)  \right].
\end{split}
\end{equation}

For nearly circular orbits, we have $r\sim a$. Hence, the second term of this equation is a constant along the orbit of the companion, being thus irrelevant to the ellipsoidal variability. However, we warn that \textbf{this term should be taken into account in non-circular orbits}.
The simple approximation when $e\sim 0$ can then be written as

\begin{equation}
\label{eq:ellip_simple}
\left(  \frac{\Delta F}{F} \right)_{\rm ell} \approx - \alpha_e ~ \frac{M_p}{M_{\star}} ~ \left( \frac{R_{\star}}{a} \right)^3 ~  \sin^2{i}\  \boldsymbol{ \cos{\left( 2\theta \right)}} ~\equiv~ -A_{\rm ell}~ \boldsymbol{ \cos{\left(2\theta \right)}},
\end{equation}

\noindent where the term $\alpha_e$ accounts for the dependence on the limb darkening and other constants.
This value is usually written in the form proposed by \cite{morris93}

\begin{equation}
\label{eq:alphae}
\alpha_e  =  0.15 \frac{ (15+\gamma)(1+g)}{3-\gamma},
\end{equation}

\noindent with $g$ being the gravity darkening coefficient.
Both $\gamma$ and $g$ values can be estimated for the specific band pass (e.g., TESS or \emph{Kepler}) from a trilinear interpolation of the effective temperature, surface gravity, and metallicity using for instance the tabulated values by \cite{claret11}.

\begin{table}[p]
\centering\small
\setlength{\tabcolsep}{5pt}
\renewcommand{\arraystretch}{1.7}
\caption{\label{tab:ellipsoidal} Ellipsoidal variations. Equations adopted by different authors.}
\begin{adjustbox}{max width=\linewidth}
\begin{tabular}{llll}

\hline  \hline

\centering   {\bf Amplitude} & \multirow{1.5}{*}{\bf Phase dependency}  & \multirow{1.5}{*}{\bf Other Factors} & \multirow{1.5}{*}{\bf Ref.$^a$} \\  [-1em]
($\Delta F_{\rm ell}/F$) & & & \\
\hline

 $-\alpha_e\frac{M_p}{M_{\star}} ~ \left( \frac{R_{\star}}{a} \right)^3 \sin^2{i}$  &  $\begin{array}{l} \cos{\left( 2\cdot 2\pi\phi \right)  } \end{array}$  &   -- & Fai11$^b$ \\

 $\beta\frac{M_p}{M_{\star}} ~ \left( \frac{R_{\star}}{r} \right)^3 \sin^3{i}$ &   $ \begin{array}{l} |\sin{\theta}| \end{array}$    &  $\beta^c=\frac{\log{(GM_{\star}/R_{\star}^2)}}{\log{T_{\rm eff}}}$ & Mis12  \\

  $-\alpha_e\frac{M_p}{M_{\star}} ~ \left( \frac{R_{\star}}{a} \right)^3 \sin^2{i}$  &   $ \begin{array}{l} \cos{(2\cdot 2\pi\phi-l)} \end{array}$  &  $l\equiv$ time lag & Bar12$^b$ \\
  $\beta\frac{M_p}{M_{\star}} ~ \left( \frac{R_{\star}}{r} \right)^3$  & $\renewcommand{\arraystretch}{1.0} \begin{array}{l}
 [\cos^2{(\omega+\nu)}+ \\ +\sin^2{(\omega+\nu)}\cos^2{i}]^{1/2}  \end{array}$  & $\beta^c$ & Kan12             \\

 $-\alpha_e\frac{M_p}{M_{\star}} ~ \left( \frac{R_{\star}}{a} \right)^3 \sin^2{i} $  &  $\renewcommand{\arraystretch}{1.0} \begin{array}{l} [\cos{(2\cdot 2\pi\phi)}+ \\ +f_1\cos{(2\pi\phi)}+ \\ +f_2\cos{(3\cdot 2\pi\phi)}]  \end{array}$      &  $f_1$,$f_2^d$  & Est13$^b$ \\
  $- \alpha_e\frac{M_p}{M_{\star}} ~ \left( \frac{R_{\star}}{a} \right)^3 \sin^2{i}$      & $ \begin{array}{l} \cos{(2\cdot 2\pi\phi)} \end{array}$     & --   & Qui13$^b$    \\
  $-\alpha_e\frac{M_p}{M_{\star}} ~ \left( \frac{R_{\star}}{r} \right)^3  \sin^2{i}$ & $ \begin{array}{l} \cos\left({2\theta}\right) \end{array}$   &  --  & Lil14 \\
\hline
\end{tabular}
\end{adjustbox}\par\medskip

\noindent\textbf{Notes:}
($^a$)~Bar12: \cite{barclay12};
Est13: \cite{esteves13};
Fai11: \cite{faigler11};
Kan12: \cite{kane12};
Lil14: \cite{lillo-box14};
Mis12: \cite{mislis12};
Qui13: \cite{quintana13}.
($^b$)~Assuming circular orbits.
($^c$)~Gravity darkening, see more in \cite{claret00a}.
($^d$)~Constants from Eq.\,9~\&~10 in their work.
\end{table}

As shown in Eq.~\ref{eq:ellip_simple}, some of the authors use other definitions of the gravity and limb darkening.
We note that \cite{mislis12} is the only work using a third-order dependence on $\sin{i}$ instead of a second-order one, which is intriguing and could potentially be a mistake in their formula.
The phase dependence, however, is more controversial, with some authors assuming circular orbits (i.e., geometric phase angle matching the temporal phase angle, $\theta(t) = 2\pi\phi(t)$) while others adopt more complex dependencies \citep[e.g., ][]{kane12, mislis12}.
We found no theoretical demonstrations or references in these works to justify such dependencies.
\cite{barclay12} introduced a time lag in the phase dependence, accounting for the possibility of a misalignment between the planet-star direction and the center of the tidal bulge.

Based on the demonstrations provided in this section, we recommend using Eq.~\ref{eq:ellip_simple} for systems with small eccentricities and Eq.~\ref{eq:ellip_2nd} for blind surveys or known high-eccentricity orbits.
Note that when additional harmonics are relevant, they will appear in the residuals of the model fit as additional modulations \citep[see, for instance the case of KOI-13.01 in ][]{mazeh12}.

\subsection{Doppler beaming}

\label{sec:beaming}

Orbiting companions induce a reflex motion of the host star around the center of masses of the system.
This motion causes small shifts in the stellar spectrum: towards longer wavelengths (redshift) when the star moves away from the observer, and towards shorter wavelengths (blueshift) when it approaches.
This phenomenon, known as the Doppler effect, is widely exploited to measure the radial velocity, and thus the mass, of planetary and multiple-star systems.
When photometry is obtained on a fixed wavelength range (using a photometric band), this effect also leaves an imprint on the measured flux \citep{Hills74}.
As the companion orbits, not only the spectral lines but the entire spectral energy distribution of the star shifts, causing the star to appear redder or bluer.
Consequently, the amount of energy received within the bandpass varies in synchrony with the orbital periodicity of the companion.

\cite{loeb03} studied this effect and derived a formulation for non-relativistic motion of the central star.
According to this work, a star moving with a radial velocity ${\rm v}_r$ relative to the observer experiences a Doppler shift in its bolometric flux given by $F=F_0(1+4\cdot {\rm v}_r/c)$, where $F_0$ is the flux in the absence of motion.
Assuming a power-law dependence of the observed light on frequency ($f$\footnote{Light frequency is typically represented by $\nu$. However, in this context, we use $f$ instead to distinguish it from the true anomaly.}) as $F_{f,0}\propto~f^{\Gamma}$, the measured flux at a specific $f$ is described by

\begin{equation}
\label{eq:beaming_tmp}
F_{f} = F_{f,0} \left[  1+(3-\Gamma)\frac{{\rm v}_r}{c}  \right]  ~  \Longrightarrow  ~ \left( \frac{\Delta F}{F} \right)_{\rm beam} = \left( 3-\Gamma \right) \frac{{\rm v}_r}{c}.
\end{equation}

The radial velocity of a star around the center of masses of the system is given by

\begin{equation}
\label{eq:RV_tmp}
{\rm v}_r(t) = V_{\rm sys} + K \left[  \cos{(\nu (t)+\omega)} +e~\cos{\omega}   \right],
\end{equation}

where $V_{\rm sys}$ is the systemic velocity and $K$ is the radial velocity semi-amplitude, given by

\begin{equation}
\label{eq:Kbeaming}
K = 28.4\,{\rm m\,s}^{-1} \times \left(\frac{P}{1 \rm yr} \right)^{-1/3} \frac{M_p\sin{i}}{M_{\rm Jup}}  \left(\frac{M_{\star}}{M_{\odot}}\right)^{-2/3} \frac{1}{\sqrt{1-e^2}}.
\end{equation}

Given that $\nu(t) +\omega -\pi/2= \theta(t)$, and $K\,e\,\cos{\omega}$ in Eq.~\ref{eq:RV_tmp} remains constant under the assumption that the orbit is neither precessing nor circularizing (i.e., $e$ and $\omega$ are not time dependent), the radial velocity of the star as a function of time can then be rewritten as

\begin{equation}
\label{eq:RV}
{\rm v}_r(t) =  K  \sin{\theta} + C.
\end{equation}

Thus, neglecting the constant term ($C$), the beaming effect is described as

 \begin{equation}
\label{eq:beaming}
\left( \frac{\Delta F}{F} \right)_{\rm beam} =   \left( 3-\Gamma \right)\frac{K}{c}  \sin{\theta}.
\end{equation}

The coefficient $\Gamma$ can be approximated as $\Gamma = d\,{\rm ln}\,F_{f 0}/d\,{\rm ln}\,f_0$. If we assume the star as a blackbody with temperature $T_{\rm eff}$, and setting $x=hf / k T_{\rm eff}$, the coefficient is

\begin{equation}
\label{eq:Gamma}
\Gamma(f) \approx \frac{{\rm e}^x(3-x)-3}{{\rm e}^x-1}.
\end{equation}

A similar but more general expression was provided by \cite{bloemen11}, who formulated a complete expression to accurately calculate the beaming effect as

\begin{equation}
\label{eq:beaming_complete}
F_{\lambda} = F_{\lambda,0} \left[  1+B\frac{{\rm v}_r}{c}  \right]  \longrightarrow \left( \frac{\Delta F}{F} \right)_{\rm beam} =  B ~\frac{K}{c}  \sin{\theta} ~ \equiv~ A_{\rm beam} ~\sin{\theta},
\end{equation}

\noindent with $B$ being the beaming factor defined as $B=5+d\,{\rm ln}\,F_{\lambda}/d\, {\rm ln}\,\lambda$; with $\lambda$ representing the wavelength.
In the case of the \emph{Kepler} observations, a photon-weighted bandpass-integrated beaming factor is used, defined as

\begin{equation}
\label{eq:beaming_factor}
<B>\,= \frac{\int \epsilon_{\lambda} \lambda F_{\lambda} B d\lambda  }{\int \epsilon_{\lambda} \lambda F_{\lambda}  d\lambda},
\end{equation}

\noindent where $\epsilon_{\lambda}$ is the response function of the observing bandpass.

By comparing Eq.~\ref{eq:beaming} and Eq.~\ref{eq:beaming_complete}, and defining $B_{\Gamma}\equiv (3-\Gamma)$, we can identify that $B_{\Gamma}=B$.
We now compare the beaming factors derived from these two approaches to assess their validity ranges and draw conclusions:

\begin{figure}[htbp]
\centering
\includegraphics[width=1\textwidth]{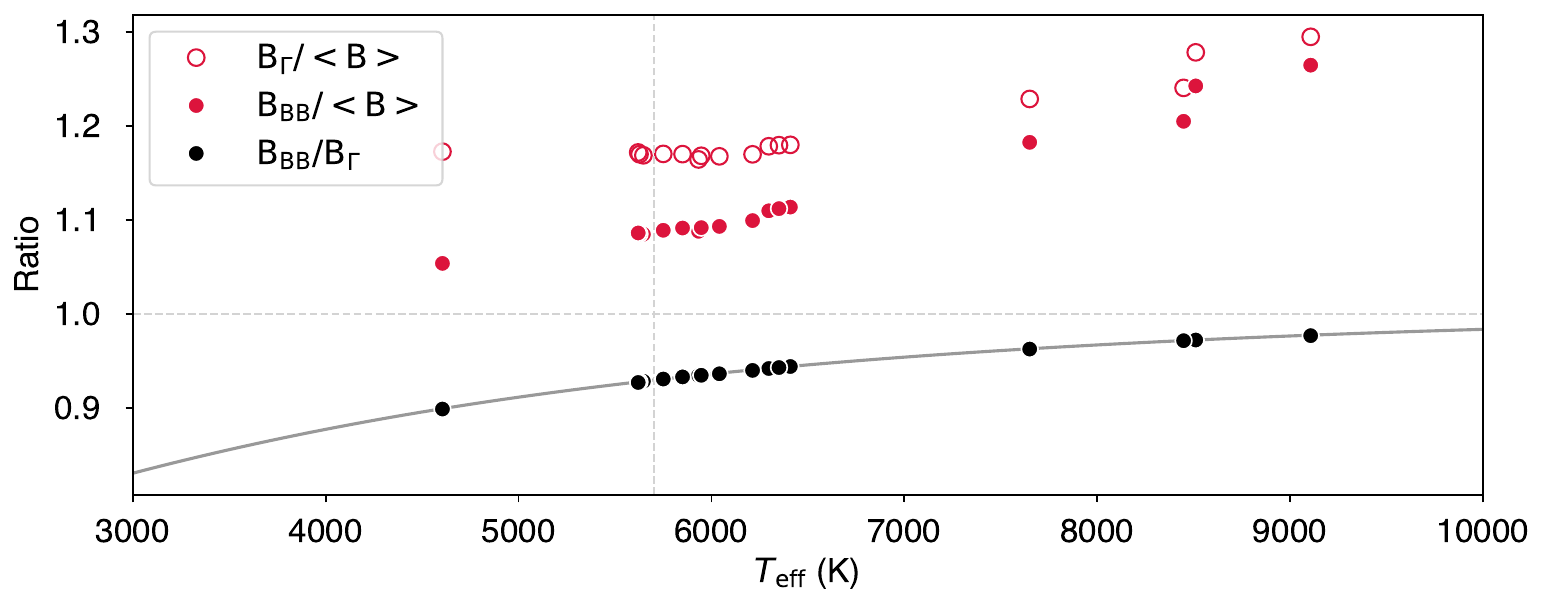}
\caption{Comparison of the different approximations of the beaming factor. Dashed vertical line shows the effective temperature of the Sun, while the horizontal line shows a ratio of one.}
\label{fig:beaming_factor}
\end{figure}

\renewcommand{\arraystretch}{1.7}
\begin{table}[p]
\centering\small
\setlength{\tabcolsep}{5pt}
\renewcommand{\arraystretch}{1.7}
\begin{center}
\caption{\label{tab:beaming} Beaming effect. Equations adopted by different authors.}
\begin{adjustbox}{max width=\linewidth}
\begin{tabular}{llll}

\hline  \hline
  {\bf Amplitude} & \multirow{1.5}{*}{\bf Phase dependency}  & \multirow{1.5}{*}{\bf Other Factors} & \multirow{1.5}{*}{\bf Ref.$^a$} \\ [-1em]
  ($\Delta F_{\rm beam}/F$) & & & \\
\hline

  $\alpha_{\rm b}\, 4\frac{K}{c}$ & $\sin{(2\pi\phi)}$   &   $\alpha_{\rm b}$ $^b$ & Fai11 \\

 $(3-\Gamma)\frac{K}{c}$ & $\sin{\theta}+e\sin{(\pi/2-\omega)}$   &  -- & Mis12 \\

 $ B\frac{K}{c}$ & $\sin{(2\pi\phi)}$  & -- & Bar12    \\

$(3-\Gamma)\frac{K}{c}$  & -- & -- & Kan12              \\

$B \frac{K}{c}$ & $\sin{(2\pi\phi)}$       & -- & Est13     \\

$\alpha_{\rm b}\, 4\frac{K}{c}$  & $\sin{(2\pi\phi)}$      &  $\alpha_{\rm b}$ $^b$ & Qui13 \\

 $(3-\Gamma)\frac{K}{c}$ & $\sin{\theta}$  & -- & Lil14 \\
\hline \\
\end{tabular}
\end{adjustbox}\par\medskip
\end{center}

\noindent\textbf{Notes:}
($^a$)~Bar12: \cite{barclay12};
Est13: \cite{esteves13};
Fai11: \cite{faigler11};
Kan12: \cite{kane12};
Lil14: \cite{lillo-box14};
Mis12: \cite{mislis12};
Qui13: \cite{quintana13}.
($^b$)~It depends on the bandpass. For \textit{Kepler} and CoRoT, it ranges from 0.8 to 1.2.
\end{table}

\begin{itemize}

\item \textbf{Blackbody approximation:} If we also consider in the general formulation from \cite{bloemen11} that the host star emits as a blackbody we would expect a similar beaming factor ($B$) than that from the \cite{loeb03} approximation ($B_{\Gamma}$).
In Fig.~\ref{fig:beaming_factor} we show this comparison
(black dots) in a range of $T_{eff} =$ 3000\,-\,10\,000\,K, where the blackbody consideration for $B$ is labeled as $B_{BB}$. We find a good agreement between these formulations (i.e., $B_{\Gamma}/B_{BB}~\approx~1$) for hotter stars.
However, there are differences of around 5-10\%.

\item \textbf{A correct beaming factor:} The most accurate method for calculating the beaming factor requires using Eq.~\ref{eq:beaming_factor}, and assuming synthetic models for the host star emission ($F_{\lambda}$) according to its physical parameters, which account for spectral lines and molecular bands effects.
In Fig.~\ref{fig:beaming_factor} we compare the three beaming factor formulations (i.e., $B_{\Gamma}$, $B_{BB}$, and $<B>$) by using the data of $<B>$ and $T_{eff}$ from \cite{esteves14}.
For the 13 planets analyzed with $T_{\rm eff}<7000~$K, we find that $B_{\Gamma}$ (Eq.~\ref{eq:Gamma}) is 17\% larger than $<B>$ (empty red circles).
The use of $B_{BB}$ (Eq.~\ref{eq:beaming_factor}) provides results closer to $<B>$ (filled red circles), but also overestimates the beaming factor by more than 10\% for this range of temperatures.
We thus strongly recommend to calculate the beaming factor as $<B>$.

\end{itemize}

Apart from the formulations described above, other authors have used different approaches as we present in Table.~\ref{tab:beaming}.
Most of the works agree in both the amplitude and the phase dependency.
The additional term in the phase dependency in the case of \cite{mislis12} can be considered constant since we work with normalized (rather than absolute) photometry.

\subsection{Degeneracies and relative contributions}
\label{sec:REBtesting}

Modeling the out-of-transit light curve of a star hosting a close-in massive planet requires the consideration of the interplay among the REB effects.
When their amplitudes are similar, they become intertwined, leading to degeneracies that difficult the derivation of the companion's properties.
In such cases, selecting suitable formalisms for the specific characteristics of the system is critical for disentangling the contributions.
In the following, we describe the most important interdependencies among these three effects:

\begin{itemize}

\item \textbf{Reflection vs. ellipsoidal:}
There is a shared dependency on $\cos{(2\theta)}$ for the ellipsoidal effect (Eq.~\ref{eq:ellip_simple}) and the second-order term of the Lambertian phase function in the reflection and thermal effects (Eq.~\ref{eq:phasefunc_LamFourier}).
In case of using the geometrical phase  function (Eq.~\ref{eq:phasefunc_geo}) instead of the Lambertian, we would be ignoring this second-order term wrongly attributing the whole $\cos{(2\theta)}$ pattern to the ellipsoidal variations.
Therefore, \textbf{a wrong phase function selection can result in a over- or under-estimation the companion mass} that is computed from the ellipsoidal contribution.
For instance, \cite{mislis12} demonstrated that the mass of HAT-P-7\,b can differ by $0.67~M_{\rm Jup}$ depending on the phase function used.
On the other hand, both Lambertian and geometrical phase functions share the $\cos{\theta}$ dependency with the third harmonic of the ellipsoidal variation ($l=3$).
\textbf{Ignoring the third harmonic in the ellipsoidal effect when relevant could lead to an underestimation of the reflection amplitude.}
However, we note that no claims have yet been made regarding the detection of the third harmonic in the ellipsoidal variations.

\item {\bf Reflection vs. beaming:}
Discrepancies in the companion mass derived from beaming and ellipsoidal modulations have been observed in the few systems where both effects were measured simultaneously.
The comparison between the RV semi-amplitude ($K$) obtained from the beaming modulation and the spectroscopically measured suggests an inflation of the beaming-derived mass.
This may be caused by a phase shift between the reflection (Eq.~\ref{eq:reflection}) and emission (Eq.~\ref{eq:thermal_day}-\ref{eq:thermal_night}) modulations due to the equatorial super-rotation of hot Jupiters, as proposed by \cite{faigler14}.
The super-rotating (SR) model, first theorized by \cite{showman02} and later confirmed by Spitzer observations (\citealt{knutson07,knutson09}), shows that tidally locked planets can develop a fast eastward jet stream from the equator to latitudes of typically $20^{\circ}-60^{\circ}$, producing a phase shift in the planet's thermal emission phase curve.
This shift introduces a new term with the same $\sin{\theta}$ dependency as the Doppler beaming effect, since $\cos{( \theta+\delta_{\rm SR})}=\cos{\delta_{\rm SR}}~\boldsymbol{\cos{\theta}}+\sin{\delta_{\rm SR}}~\boldsymbol{\sin{\theta}}$, with $\delta_{\rm SR}$ being the angle shift.
Thus, in a \textbf{super-rotating planet we might over-estimate both the beaming effect and its derived mass if not including the phase shift in the reflection effect}.
In case of using a Lambertian phase function, this shift introduces an additional $\sin{2\theta}$ dependency as described in \cite{faigler14}. Hence, to account for the SR effect, the phase functions take the form

\begin{align}
\begin{split}
\Phi_{\rm geo}^{\rm SR} (z)   ~&~ =  -\sin{i}~  \Big[ \cos{\delta_{\rm SR}}~\boldsymbol{\cos{\theta}}+\sin{\delta_{\rm SR}}~\boldsymbol{\sin{\theta}} \Big] \\
\Phi_{\rm Lam}^{\rm SR} (z) ~&~ \approx   \Phi_{\rm geo}^{\rm SR}(z) + 0.18\sin^2i ~ \Big[ \cos{(2\delta_{\rm SR})} ~ \boldsymbol{\cos{(2\theta)}} + \\
~&~  \hspace{3.5cm} + \sin{(2\delta_{\rm SR})} ~\boldsymbol{\sin{(2\theta)}}\Big] .
\end{split}
\end{align}

\end{itemize}

\begin{figure}[htbp]
\centering
\includegraphics[width=1\textwidth]{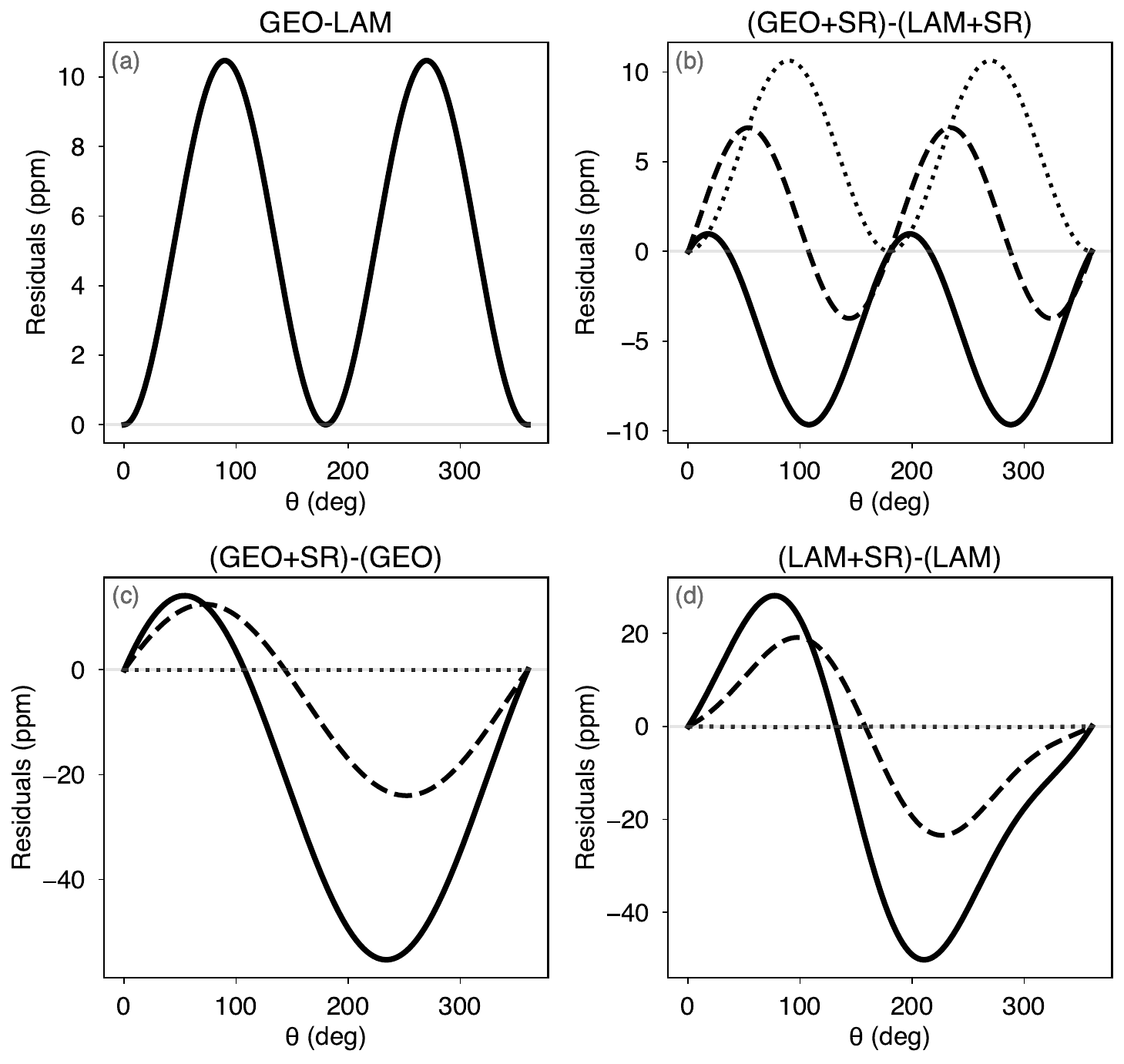}
\caption{Differences between the formulations from Table~\ref{tab:REBdependencies} for models with fixed amplitudes and orbital inclination ($A_{\rm ell}=60$~ppm, $A_{\rm ref}=30$~ppm, $A_{\rm beam}=10$~ppm, $i=80^{\circ}$). The title of each panel (a, b, c and d) describes the difference computed. We show three cases of phase shifts for the super-rotation effect: $\delta_{\rm SR}=0$ (dotted line), $\delta_{\rm SR}=0.1$ (dashed), and $\delta_{\rm SR}=0.2$ (solid), in phase units.}
\label{fig:REBmodels}
\end{figure}

Table~\ref{tab:REBdependencies} summarizes the dependencies of the different effects on the geometrical angle $\theta$, considering both the presence and absence of super-rotation, and using the two phase functions fully described in Sect.~\ref{sec:reflection}.
To explore the differences between the formulations presented in that table, we computed them for specific amplitudes and inclination ($A_{\rm ell}=60$~ppm, $A_{\rm ref}=30$~ppm, $A_{\rm beam}=10$~ppm, $i=80^{\circ}$).
Fig.~\ref{fig:REBmodels} shows the residuals between different pairs of formulations, assuming various phase shifts for the super-rotation effect ($\delta_{\rm SR}=0, 0.1, 0.2$).
Panel (a) illustrates that the residuals between the Lambertian and the geometrical phase functions exhibit the same shape as the ellipsoidal effect, as discussed above.
In panel (b), the inclusion of super-rotation introduces a phase shift in the residuals, but the amplitudes remain unchanged.
Panels (c) and (d) show the inflation of the beaming effect when the super-rotation is neglected for each phase function.

\begin{table}[p]
\centering\small
\setlength{\tabcolsep}{5pt}
\renewcommand{\arraystretch}{1.7}
\caption{\label{tab:REBdependencies} Amplitudes of the different angle dependencies for the REB modulations.}
\begin{adjustbox}{max width=\linewidth}
\begin{tabular}{l|ll|ll}
\hline  \hline

$\boldsymbol{\Phi(z)}$  & \multicolumn{2}{l|}{Geometrical} & \multicolumn{2}{l}{Lambertian} \\ [-0.5em]
\textbf{SR}$^a$ & No & Yes & No & Yes \\
\hline
$\boldsymbol{\cos{\theta}}$ & $-A_{\rm ref}$ & $-A_{\rm ref}\cos{\delta_{\rm SR}}$ & $-A_{\rm ref}$ & $-A_{\rm ref}\cos{\delta_{\rm SR}}$ \\
$\boldsymbol{\sin{\theta}}$ & $A_{\rm beam}$ & $A_{\rm beam}+A_{\rm ref}\sin{\delta_{\rm SR}}$ & $A_{\rm beam}$ & $A_{\rm beam}+A_{\rm ref}\sin{\delta_{\rm SR}}$ \\
$\boldsymbol{\cos{\left( 2 \theta \right)}}$ & $-A_{\rm ell}$ & $-A_{\rm ell}$ & $\renewcommand{\arraystretch}{1.0}  \begin{array}{l} -A_{\rm ell}+ \\+0.18~A_{\rm ref}~\sin{i} \end{array}$ & $ \renewcommand{\arraystretch}{1.0} \begin{array}{l}
-A_{\rm ell}+0.18~A_{\rm ref} \\ \sin{i}\cos{2\delta_{\rm SR}} \end{array}$  \\
$\boldsymbol{\sin{\left( 2 \theta \right)}}$  & 0 & 0 & 0 & $\renewcommand{\arraystretch}{1.0}  \begin{array}{l} -0.18A_{\rm ref} \\ \sin{i}\sin{2\delta_{\rm SR}} \end{array}$ \\

\hline \noalign{\smallskip}
\end{tabular}
\end{adjustbox}\par\medskip

\noindent\textbf{Notes:}
Adapted from Table 1 in \citep{faigler14}. We have only taken into account the second harmonic for the ellipsoidal effect.
($^a$)~Super-rotation.
\end{table}

\section{Demographics of the detected planets}

The REB modulations provide a unique window into a distinct subset of the planetary population. By analyzing the amplitudes of these photometric effects, we can identify the type of exoplanets that can be accessed through this technique. By neglecting constant terms, focusing on the first harmonics, and considering circular orbits ($r = a$), we find from equations in Sect.~\ref{sec:method} that

\begin{equation}
\begin{split}
	A_{\rm ell} \propto ~  & ~ \frac{M_p}{M_{\star}} ~ \left( \frac{R_{\star}}{a} \right)^3 ~  \sin^2{i} \\
	A_{\rm ref} \propto & ~ \left(\frac{R_p}{a}\right)^2 \\
	A_{\rm beam} \propto & ~
    \frac{M_p}{M_{\star}^{1/2}} \times \frac{1}{a^{1/2}} \times \sin{i}. \\
\end{split}
\end{equation}

Larger planets, in both mass and radius, induce stronger signals for all three effects.
Additionally, the two gravitationally driven effects, ellipsoidal and beaming, are also enhanced when orbiting lower-mass host stars.
Shorter orbital periods further amplify these modulations.
However, the strength of each effect scales differently with the orbital separation: ellipsoidal variations scale as $a^{-3}$, reflection as $a^{-2}$, and beaming as $a^{-1/2}$.
This means that as we move towards closer orbits, ellipsoidal variations dominate, particularly as the planet's orbit approaches the stellar radius.
This steep dependency makes ellipsoidal modulations especially significant for close-in planets orbiting giant stars, where the large stellar radius amplifies the effect.
Although such systems are relatively rare, this population grew during the 2010s with the Kepler mission (e.g., \citealt{faigler13,lillo-box14,millholland17,lillo-box21}).

Within main-sequence stars, earlier spectral types drastically reduce the amplitude of the ellipsoidal and beaming modulations as a result of their higher stellar masses.
Nonetheless, their larger temperatures and radii boost the planetary reflection.
Conversely, late-type stars allow for stronger ellipsoidal and beaming signals.

\begin{figure}[htbp]
\centering
\includegraphics[width=\textwidth]{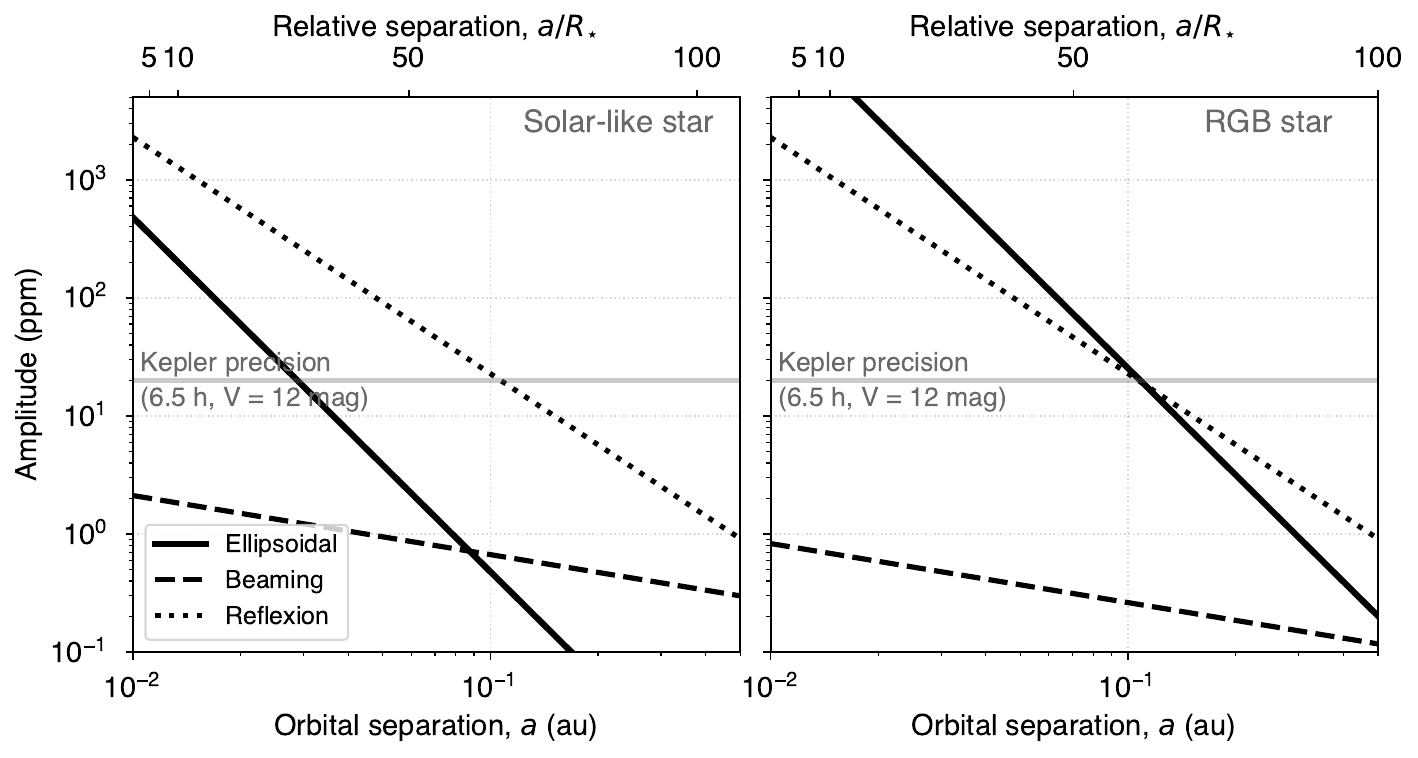}
\caption{Amplitude of the three main effects (ellipsoidal, beaming and reflection) of the orbital phase modulations depending on the orbital semi-major axis for a Jupiter-like planet ($M_p=1M_{\rm Jup}$, $R_p=1R_{\rm Jup}$) around a solar-like main-sequence star (left panel) and a star ascending the Red Giant Branch with $M_{\star}=1.3M_{\odot}$, $R_{\star}=7R_{\odot}$ (right panel). For reference, the typical Kepler combined differential photometric precision (CDPP) of 20 ppm over 6.5h is marked as an horizontal gray line (similar precision is expected for the PLATO mission).}
\label{fig:REBamplitudes}
\end{figure}

To test the impact of stellar evolution we compare two scenarios: the same Jupiter-like planet orbiting both a solar-like star and a Red Giant Branch (RGB) star with $M_{\star}=1.3M_{\odot}$ and $R_{\star}=7R_{\odot}$.
Figure~\ref{fig:REBamplitudes} illustrates the results for the three effects in these two cases as a function of the orbital separation.
The ellipsoidal modulations are notably amplified from the main sequence to the RGB phase, as dictated by their $R_{\star}^3$ dependency.
Hence, for a given photometric precision we can detect planets at longer-period orbits around RGB stars. For the same reason, we will not be able to use the technique around late evolutionary stages like in the white dwarf somain, as their radii are extremely small, typically comparable to that of Earth, making the method ineffective in these stars.

In terms of the orbital inclination, the beaming effect scales as $\sin{i}$, while the ellipsoidal variations follow a $\sin^2{i}$ dependency.
Such difference provides a potential means to determine the inclination of non-transiting planets if both effects are detected with sufficient precision, thereby breaking the degeneracy associated with this parameter. In this sense, reflected light also offers a way to estimate the orbital inclination through high-resolution spectroscopy \citep{brogi12}.

The sample of confirmed planets detected through orbital phase modulations remains quite limited, with only nine reported in the NASA Exoplanet Archive as of 2024. Five of which are considered controversial, while the remaining four are KIC\,10068024\,b, KIC\,5479689\,b, KIC\,8121913\,b (all non-transiting and detected by \citealt{lillo-box21}), and Kepler-76\,b (\citealt{faigler13}) whose detection was driven by its transit.
The main reason is that this method has been mostly used as a characterization technique for transiting planets.
For instance, the BEER program \citep{faigler12} used Kepler and CoRoT light curves to look for these effects in the out-of-transit data from known transiting candidates.
While the detection of new non-transiting planets via the REB method is possible, it is a challenging task.
The shallow amplitudes predominantly with sinusoidal shapes, as we have seen along this section, can be mimicked by many other effects, including stellar activity, pulsations, and systematics.
Still, some authors have carried out searches for non-transiting planets, and also low-mass stellar companions, with this method (e.g., \citealt{millholland17,cullen24, aller20, aller24}), producing lists of planet candidates that can later be confirmed through other techniques (e.g., \citealt{lillo-box21}).

\section{Example analysis of a real dataset}

\begin{figure}[htbp]
\centering
\includegraphics[width=1\textwidth]{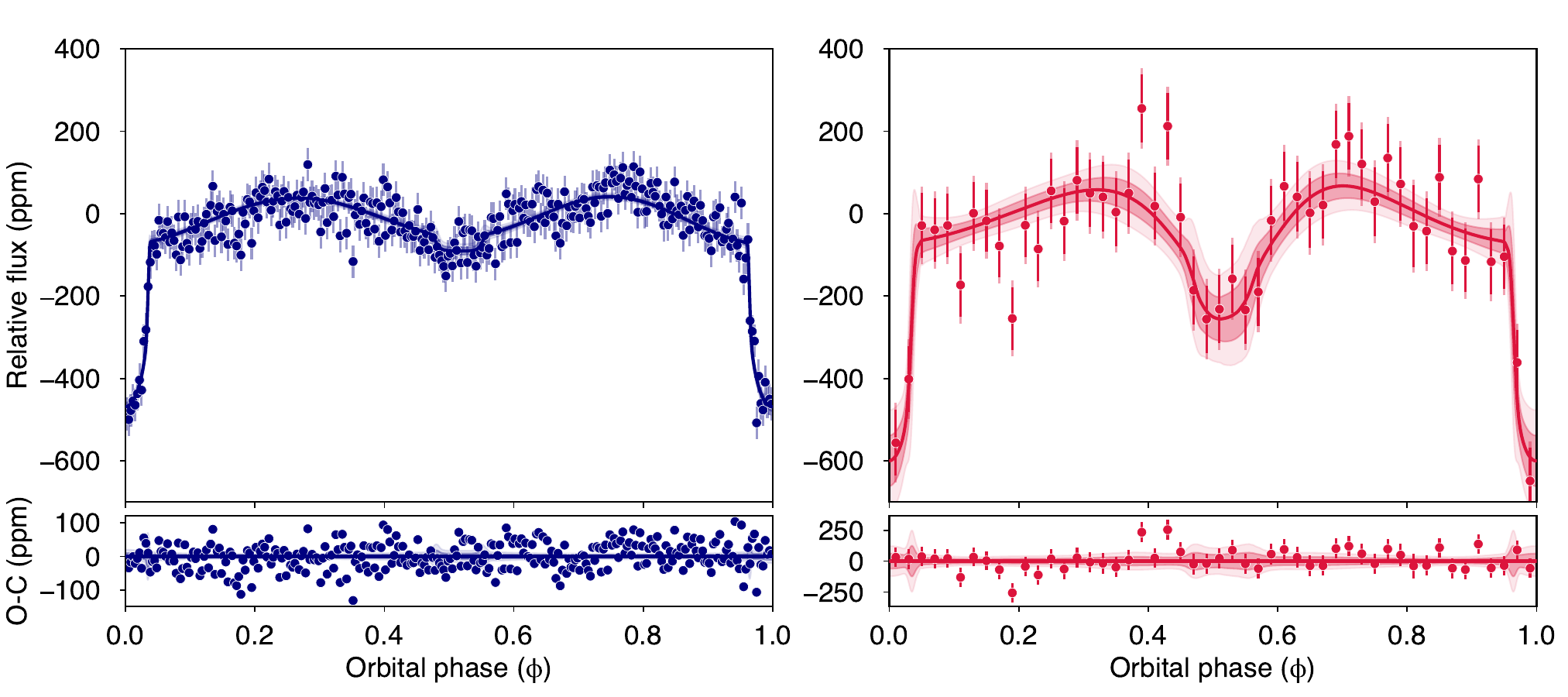}
\caption{REB modulations in the out-of-transit light curve using data from the Kepler (left panel) and TESS missions (right panel). }
\label{fig:Kepler91}
\end{figure}

Kepler-91\,b was one of the first extrasolar planets discovered to exhibit ellipsoidal modulations \citep{lillo-box14}.
The REB modulations were detected by both the Kepler and TESS missions (see Figure~\ref{fig:Kepler91}).
For this example, we only focus on the more precise Kepler dataset.
Since this is a transiting exoplanet, we leverage the known orbital period ($P_{\rm orb}=6.246696\pm0.000040$ days, \citealt{barclay14}) and mid-transit time ($T_0=2454969.3837 \pm 0.0042$, \citealt{barclay14}) only with the purpose of saving computational time.
In this example, we provide a phase-folded version of the light curve, which is subsequently binned with a size corresponding to 1/1000$^{th}$ of the orbital phase.
Additionally, we mask out the transit times to simplify the analysis by avoiding the complexities introduced by the transit signal.

For the modeling process we use the \texttt{rebeal}\footnote{\url{https://github.com/jlillo/rebeal}} software, based on the equations presented in this article and in \cite{lillo-box14}. %The material (data and code) is presented in the form of a Python Notebook that runs in Python 3.11 and requires a minimum set of parameters.

There are several libraries and softwares available to model the REB variations. Table~\ref{tab:REBsoftware} lists some of the most well-known tools to model those modulations through different approaches, from pure optimization to Monte Carlo techniques.

\begin{center}\begin{minipage}{\linewidth}\captionsetup{type=table}
\centering\small
\setlength{\tabcolsep}{5pt}
\renewcommand{\arraystretch}{1.7}
\caption{Non-exhaustive list of software tools to model REB modulations for planetary systems.}
\begin{center}
\begin{adjustbox}{max width=\linewidth}
\begin{tabular}{lll}
 \hline \hline
 Software$^a$ & Language & Reference  \\ [0.5ex]
 \hline
\href{https://github.com/MNGuenther/allesfitter}{\texttt{allesfitter}} & Python &  \cite{gunther20} \\
 \href{https://phoebe-project.org}{\texttt{phoebe}}      &  C++ (with Python interpreter) &  \cite{phoebe} \\
 \href{https://github.com/zpenoyre/OoT/wiki}{\texttt{OoT}}      & Python  &  \cite{penoyre18} \\
 \href{https://www.astro.keele.ac.uk/jkt/codes/jktebop.html}{\texttt{JKTEBOP}}      &  IDL &  \cite{southworth08} \\
 \href{https://github.com/jlillo/rebeal}{\texttt{rebeal}}       &  Python &  \cite{lillo-box14} \\
 \hline
\end{tabular}
\end{adjustbox}\par\medskip
\vspace{0.5em}
\noindent\textbf{Notes:}
($^a$)~Code names link to their websites where a URL is available.
\label{tab:REBsoftware}
\end{center}
\end{minipage}\end{center}

\par\bigskip
\bibliographystyle{plainnat}
\bibliography{references}

@article{Hills74,
	adsurl = {https://ui.adsabs.harvard.edu/abs/1974A&A....30..135H},
	author = {{Hills}, J.~G. and {Dale}, T.~M.},
	journal = {\aap},
	month = jan,
	pages = {135-139},
	title = {{The orbit evolution of the eclipsing binary system BD +16 516 and the rotation period of its white dwarf.}},
	volume = {30},
	year = 1974}

@article{aller20,
	adsurl = {https://ui.adsabs.harvard.edu/abs/2020A&A...635A.128A},
	archiveprefix = {arXiv},
	author = {{Aller}, A. and {Lillo-Box}, J. and {Jones}, D. and {Miranda}, L.~F. and {Barcel{\'o} Forteza}, S.},
	doi = {10.1051/0004-6361/201937118},
	eid = {A128},
	eprint = {1911.09991},
	journal = {\aap},
	month = mar,
	pages = {A128},
	primaryclass = {astro-ph.SR},
	title = {{Planetary nebulae seen with TESS: Discovery of new binary central star candidates from Cycle 1}},
	volume = {635},
	year = 2020}

@article{aller24,
	adsurl = {https://ui.adsabs.harvard.edu/abs/2024A&A...690A.190A},
	archiveprefix = {arXiv},
	author = {{Aller}, Alba and {Lillo-Box}, Jorge and {Jones}, David},
	doi = {10.1051/0004-6361/202450942},
	eid = {A190},
	eprint = {2409.06332},
	journal = {\aap},
	month = oct,
	pages = {A190},
	primaryclass = {astro-ph.SR},
	title = {{Planetary nebulae seen with TESS: New and revisited short-period binary central star candidates from Cycles 1 to 4}},
	volume = {690},
	year = 2024}

@article{auvergne09,
	adsurl = {https://ui.adsabs.harvard.edu/abs/2009A&A...506..411A},
	archiveprefix = {arXiv},
	author = {{Auvergne}, M. and {Bodin}, P. and {Boisnard}, L. and {Buey}, J. -T. and {Chaintreuil}, S. and {Epstein}, G. and {Jouret}, M. and {Lam-Trong}, T. and {Levacher}, P. and {Magnan}, A. and {Perez}, R. and {Plasson}, P. and {Plesseria}, J. and {Peter}, G. and {Steller}, M. and {Tiph{\`e}ne}, D. and {Baglin}, A. and {Agogu{\'e}}, P. and {Appourchaux}, T. and {Barbet}, D. and {Beaufort}, T. and {Bellenger}, R. and {Berlin}, R. and {Bernardi}, P. and {Blouin}, D. and {Boumier}, P. and {Bonneau}, F. and {Briet}, R. and {Butler}, B. and {Cautain}, R. and {Chiavassa}, F. and {Costes}, V. and {Cuvilho}, J. and {Cunha-Parro}, V. and {de Oliveira Fialho}, F. and {Decaudin}, M. and {Defise}, J. -M. and {Djalal}, S. and {Docclo}, A. and {Drummond}, R. and {Dupuis}, O. and {Exil}, G. and {Faur{\'e}}, C. and {Gaboriaud}, A. and {Gamet}, P. and {Gavalda}, P. and {Grolleau}, E. and {Gueguen}, L. and {Guivarc'h}, V. and {Guterman}, P. and {Hasiba}, J. and {Huntzinger}, G. and {Hustaix}, H. and {Imbert}, C. and {Jeanville}, G. and {Johlander}, B. and {Jorda}, L. and {Journoud}, P. and {Karioty}, F. and {Kerjean}, L. and {Lafond}, L. and {Lapeyrere}, V. and {Landiech}, P. and {Larqu{\'e}}, T. and {Laudet}, P. and {Le Merrer}, J. and {Leporati}, L. and {Leruyet}, B. and {Levieuge}, B. and {Llebaria}, A. and {Martin}, L. and {Mazy}, E. and {Mesnager}, J. -M. and {Michel}, J. -P. and {Moalic}, J. -P. and {Monjoin}, W. and {Naudet}, D. and {Neukirchner}, S. and {Nguyen-Kim}, K. and {Ollivier}, M. and {Orcesi}, J. -L. and {Ottacher}, H. and {Oulali}, A. and {Parisot}, J. and {Perruchot}, S. and {Piacentino}, A. and {Pinheiro da Silva}, L. and {Platzer}, J. and {Pontet}, B. and {Pradines}, A. and {Quentin}, C. and {Rohbeck}, U. and {Rolland}, G. and {Rollenhagen}, F. and {Romagnan}, R. and {Russ}, N. and {Samadi}, R. and {Schmidt}, R. and {Schwartz}, N. and {Sebbag}, I. and {Smit}, H. and {Sunter}, W. and {Tello}, M. and {Toulouse}, P. and {Ulmer}, B. and {Vandermarcq}, O. and {Vergnault}, E. and {Wallner}, R. and {Waultier}, G. and {Zanatta}, P.},
	doi = {10.1051/0004-6361/200810860},
	eprint = {0901.2206},
	journal = {\aap},
	month = oct,
	number = {1},
	pages = {411-424},
	primaryclass = {astro-ph.SR},
	title = {{The CoRoT satellite in flight: description and performance}},
	volume = {506},
	year = 2009}

@ARTICLE{barbier21,
       author = {{Barbier}, H. and {L{\'o}pez}, E.},
        title = "{Kepler Planetary Systems: Doppler Beaming Effect Significance}",
      journal = {\rmxaa},
         year = 2021,
        month = apr,
       volume = {57},
        pages = {123-132},
          doi = {10.22201/ia.01851101p.2021.57.01.08},
       adsurl = {https://ui.adsabs.harvard.edu/abs/2021RMxAA..57..123B}
}

@article{barclay12,
	adsurl = {http://adsabs.harvard.edu/abs/2012ApJ...761...53B},
	archiveprefix = {arXiv},
	author = {{Barclay}, T. and {Huber}, D. and {Rowe}, J.~F. and {Fortney}, J.~J. and {Morley}, C.~V. and {Quintana}, E.~V. and {Fabrycky}, D.~C. and {Barentsen}, G. and {Bloemen}, S. and {Christiansen}, J.~L. and {Demory}, B.-O. and {Fulton}, B.~J. and {Jenkins}, J.~M. and {Mullally}, F. and {Ragozzine}, D. and {Seader}, S.~E. and {Shporer}, A. and {Tenenbaum}, P. and {Thompson}, S.~E.},
	doi = {10.1088/0004-637X/761/1/53},
	eid = {53},
	eprint = {1210.4592},
	journal = {\apj},
	month = dec,
	pages = {53},
	primaryclass = {astro-ph.SR},
	title = {{Photometrically Derived Masses and Radii of the Planet and Star in the TrES-2 System}},
	volume = 761,
	year = 2012}

@article{barclay14,
	adsurl = {http://adsabs.harvard.edu/abs/2015ApJ...800...46B},
	archiveprefix = {arXiv},
	author = {{Barclay}, T. and {Endl}, M. and {Huber}, D. and {Foreman-Mackey}, D. and {Cochran}, W.~D. and {MacQueen}, P.~J. and {Rowe}, J.~F. and {Quintana}, E.~V.},
	doi = {10.1088/0004-637X/800/1/46},
	eid = {46},
	eprint = {1408.3149},
	journal = {\apj},
	month = feb,
	pages = {46},
	primaryclass = {astro-ph.EP},
	title = {{Radial Velocity Observations and Light Curve Noise Modeling Confirm that Kepler-91b is a Giant Planet Orbiting a Giant Star}},
	volume = 800,
	year = 2015}

@ARTICLE{barros22,
       author = {{Barros}, S.~C.~C. and {Akinsanmi}, B. and {Bou{\'e}}, G. and {Smith}, A.~M.~S. and {Laskar}, J. and {Ulmer-Moll}, S. and {Lillo-Box}, J. and {Queloz}, D. and {Cameron}, A. Collier and {Sousa}, S.~G. and {Ehrenreich}, D. and {Hooton}, M.~J. and {Bruno}, G. and {Demory}, B. -O. and {Correia}, A.~C.~M. and {Demangeon}, O.~D.~S. and {Wilson}, T.~G. and {Bonfanti}, A. and {Hoyer}, S. and {Alibert}, Y. and {Alonso}, R. and {Escud{\'e}}, G. Anglada and {Barbato}, D. and {B{\'a}rczy}, T. and {Barrado}, D. and {Baumjohann}, W. and {Beck}, M. and {Beck}, T. and {Benz}, W. and {Bergomi}, M. and {Billot}, N. and {Bonfils}, X. and {Bouchy}, F. and {Brandeker}, A. and {Broeg}, C. and {Cabrera}, J. and {Cessa}, V. and {Charnoz}, S. and {Damme}, C.~C.~V. and {Davies}, M.~B. and {Deleuil}, M. and {Deline}, A. and {Delrez}, L. and {Erikson}, A. and {Fortier}, A. and {Fossati}, L. and {Fridlund}, M. and {Gandolfi}, D. and {Mu{\~n}oz}, A. Garc{\'\i}a and {Gillon}, M. and {G{\"u}del}, M. and {Isaak}, K.~G. and {Heng}, K. and {Kiss}, L. and {des Etangs}, A. Lecavelier and {Lendl}, M. and {Lovis}, C. and {Magrin}, D. and {Nascimbeni}, V. and {Maxted}, P.~F.~L. and {Olofsson}, G. and {Ottensamer}, R. and {Pagano}, I. and {Pall{\'e}}, E. and {Parviainen}, H. and {Peter}, G. and {Piotto}, G. and {Pollacco}, D. and {Ragazzoni}, R. and {Rando}, N. and {Rauer}, H. and {Ribas}, I. and {Santos}, N.~C. and {Scandariato}, G. and {S{\'e}gransan}, D. and {Simon}, A.~E. and {Steller}, M. and {Szab{\'o}}, Gy. M. and {Thomas}, N. and {Udry}, S. and {Ulmer}, B. and {Van Grootel}, V. and {Walton}, N.~A.},
        title = "{Detection of the tidal deformation of WASP-103b at 3 {\ensuremath{\sigma}} with CHEOPS}",
      journal = {\aap},
         year = 2022,
        month = jan,
       volume = {657},
          eid = {A52},
        pages = {A52},
          doi = {10.1051/0004-6361/202142196},
archivePrefix = {arXiv},
       eprint = {2201.03328},
 primaryClass = {astro-ph.EP},
       adsurl = {https://ui.adsabs.harvard.edu/abs/2022A&A...657A..52B}
}

@article{bloemen11,
	adsurl = {http://adsabs.harvard.edu/abs/2011MNRAS.410.1787B},
	archiveprefix = {arXiv},
	author = {{Bloemen}, S. and {Marsh}, T.~R. and {{\O}stensen}, R.~H. and {Charpinet}, S. and {Fontaine}, G. and {Degroote}, P. and {Heber}, U. and {Kawaler}, S.~D. and {Aerts}, C. and {Green}, E.~M. and {Telting}, J. and {Brassard}, P. and {G{\"a}nsicke}, B.~T. and {Handler}, G. and {Kurtz}, D.~W. and {Silvotti}, R. and {Van Grootel}, V. and {Lindberg}, J.~E. and {Pursimo}, T. and {Wilson}, P.~A. and {Gilliland}, R.~L. and {Kjeldsen}, H. and {Christensen-Dalsgaard}, J. and {Borucki}, W.~J. and {Koch}, D. and {Jenkins}, J.~M. and {Klaus}, T.~C.},
	doi = {10.1111/j.1365-2966.2010.17559.x},
	eprint = {1010.2747},
	journal = {\mnras},
	month = jan,
	pages = {1787-1796},
	primaryclass = {astro-ph.SR},
	title = {{Kepler observations of the beaming binary KPD 1946+4340}},
	volume = 410,
	year = 2011}

@inproceedings{bond00,
	adsurl = {https://ui.adsabs.harvard.edu/abs/2000ASPC..199..115B},
	archiveprefix = {arXiv},
	author = {{Bond}, Howard E.},
	booktitle = {Asymmetrical Planetary Nebulae II: From Origins to Microstructures},
	doi = {10.48550/arXiv.astro-ph/9909516},
	editor = {{Kastner}, J.~H. and {Soker}, N. and {Rappaport}, S.},
	eprint = {astro-ph/9909516},
	month = jan,
	pages = {115},
	primaryclass = {astro-ph},
	series = {Astronomical Society of the Pacific Conference Series},
	title = {{Binarity of Central Stars of Planetary Nebulae}},
	volume = {199},
	year = 2000}

@article{borucki09,
	adsurl = {http://adsabs.harvard.edu/abs/2009Sci...325..709B},
	author = {{Borucki}, W.~J. and {Koch}, D. and {Jenkins}, J. and {Sasselov}, D. and {Gilliland}, R. and {Batalha}, N. and {Latham}, D.~W. and {Caldwell}, D. and {Basri}, G. and {Brown}, T. and {Christensen-Dalsgaard}, J. and {Cochran}, W.~D. and {DeVore}, E. and {Dunham}, E. and {Dupree}, A.~K. and {Gautier}, T. and {Geary}, J. and {Gould}, A. and {Howell}, S. and {Kjeldsen}, H. and {Lissauer}, J. and {Marcy}, G. and {Meibom}, S. and {Morrison}, D. and {Tarter}, J.},
	doi = {10.1126/science.1178312},
	journal = {Science},
	month = aug,
	pages = {709-},
	title = {{Kepler{\rsquo}s Optical Phase Curve of the Exoplanet HAT-P-7b}},
	volume = 325,
	year = 2009}

@article{borucki10,
	adsurl = {http://adsabs.harvard.edu/abs/2010Sci...327..977B},
	author = {{Borucki}, W.~J. and {Koch}, D. and {Basri}, G. and {Batalha}, N. and {Brown}, T. and {Caldwell}, D. and {Caldwell}, J. and {Christensen-Dalsgaard}, J. and {Cochran}, W.~D. and {DeVore}, E. and {Dunham}, E.~W. and {Dupree}, A.~K. and {Gautier}, T.~N. and {Geary}, J.~C. and {Gilliland}, R. and {Gould}, A. and {Howell}, S.~B. and {Jenkins}, J.~M. and {Kondo}, Y. and {Latham}, D.~W. and {Marcy}, G.~W. and {Meibom}, S. and {Kjeldsen}, H. and {Lissauer}, J.~J. and {Monet}, D.~G. and {Morrison}, D. and {Sasselov}, D. and {Tarter}, J. and {Boss}, A. and {Brownlee}, D. and {Owen}, T. and {Buzasi}, D. and {Charbonneau}, D. and {Doyle}, L. and {Fortney}, J. and {Ford}, E.~B. and {Holman}, M.~J. and {Seager}, S. and {Steffen}, J.~H. and {Welsh}, W.~F. and {Rowe}, J. and {Anderson}, H. and {Buchhave}, L. and {Ciardi}, D. and {Walkowicz}, L. and {Sherry}, W. and {Horch}, E. and {Isaacson}, H. and {Everett}, M.~E. and {Fischer}, D. and {Torres}, G. and {Johnson}, J.~A. and {Endl}, M. and {MacQueen}, P. and {Bryson}, S.~T. and {Dotson}, J. and {Haas}, M. and {Kolodziejczak}, J. and {Van Cleve}, J. and {Chandrasekaran}, H. and {Twicken}, J.~D. and {Quintana}, E.~V. and {Clarke}, B.~D. and {Allen}, C. and {Li}, J. and {Wu}, H. and {Tenenbaum}, P. and {Verner}, E. and {Bruhweiler}, F. and {Barnes}, J. and {Prsa}, A.},
	doi = {10.1126/science.1185402},
	journal = {Science},
	month = feb,
	pages = {977-},
	title = {{Kepler Planet-Detection Mission: Introduction and First Results}},
	volume = 327,
	year = 2010}

@ARTICLE{brogi12,
       author = {{Brogi}, Matteo and {Snellen}, Ignas A.~G. and {de Kok}, Remco J. and {Albrecht}, Simon and {Birkby}, Jayne and {de Mooij}, Ernst J.~W.},
        title = "{The signature of orbital motion from the dayside of the planet {\ensuremath{\tau}} Bo{\"o}tis b}",
      journal = {\nat},
         year = 2012,
        month = jun,
       volume = {486},
       number = {7404},
        pages = {502-504},
          doi = {10.1038/nature11161},
archivePrefix = {arXiv},
       eprint = {1206.6109},
 primaryClass = {astro-ph.EP},
       adsurl = {https://ui.adsabs.harvard.edu/abs/2012Natur.486..502B}
}

@article{cahoy10,
	adsurl = {http://adsabs.harvard.edu/abs/2010ApJ...724..189C},
	archiveprefix = {arXiv},
	author = {{Cahoy}, K.~L. and {Marley}, M.~S. and {Fortney}, J.~J.},
	doi = {10.1088/0004-637X/724/1/189},
	eprint = {1009.3071},
	journal = {\apj},
	month = nov,
	pages = {189-214},
	primaryclass = {astro-ph.EP},
	title = {{Exoplanet Albedo Spectra and Colors as a Function of Planet Phase, Separation, and Metallicity}},
	volume = 724,
	year = 2010}

@ARTICLE{charbonneau99,
       author = {{Charbonneau}, David and {Noyes}, Robert W. and {Korzennik}, Sylvain G. and {Nisenson}, Peter and {Jha}, Saurabh and {Vogt}, Steven S. and {Kibrick}, Robert I.},
        title = "{An Upper Limit on the Reflected Light from the Planet Orbiting the Star {\ensuremath{\tau}} Bootis}",
      journal = {\apjl},
         year = 1999,
        month = sep,
       volume = {522},
       number = {2},
        pages = {L145-L148},
          doi = {10.1086/312234},
archivePrefix = {arXiv},
       eprint = {astro-ph/9907195},
 primaryClass = {astro-ph},
       adsurl = {https://ui.adsabs.harvard.edu/abs/1999ApJ...522L.145C}
}

@ARTICLE{claret00a,
       author = {{Claret}, A.},
        title = "{Studies on stellar rotation. II. Gravity-darkening: the effects of the input physics and differential rotation. New results for very low mass stars}",
      journal = {\aap},
         year = 2000,
        month = jul,
       volume = {359},
        pages = {289-298},
       adsurl = {https://ui.adsabs.harvard.edu/abs/2000A&A...359..289C}
}

@article{claret11,
	adsurl = {http://adsabs.harvard.edu/abs/2011A%26A...529A..75C},
	author = {{Claret}, A. and {Bloemen}, S.},
	doi = {10.1051/0004-6361/201116451},
	eid = {A75},
	journal = {\aap},
	month = may,
	pages = {A75},
	title = {{Gravity and limb-darkening coefficients for the Kepler, CoRoT, Spitzer, uvby, UBVRIJHK, and Sloan photometric systems}},
	volume = 529,
	year = 2011}

@ARTICLE{collier-cameron99,
       author = {{Collier Cameron}, Andrew and {Horne}, Keith and {Penny}, Alan and {James}, David},
        title = "{Probable detection of starlight reflected from the giant planet orbiting {\ensuremath{\tau}} Bo{\"o}tis}",
      journal = {\nat},
         year = 1999,
        month = dec,
       volume = {402},
       number = {6763},
        pages = {751-755},
          doi = {10.1038/45451},
archivePrefix = {arXiv},
       eprint = {astro-ph/9911314},
 primaryClass = {astro-ph},
       adsurl = {https://ui.adsabs.harvard.edu/abs/1999Natur.402..751C}
}

@article{cowan11,
	adsurl = {http://adsabs.harvard.edu/abs/2011ApJ...729...54C},
	archiveprefix = {arXiv},
	author = {{Cowan}, N.~B. and {Agol}, E.},
	doi = {10.1088/0004-637X/729/1/54},
	eid = {54},
	eprint = {1001.0012},
	journal = {\apj},
	month = mar,
	pages = {54},
	primaryclass = {astro-ph.EP},
	title = {{The Statistics of Albedo and Heat Recirculation on Hot Exoplanets}},
	volume = 729,
	year = 2011}

@ARTICLE{cowan12,
       author = {{Cowan}, Nicolas B. and {Machalek}, Pavel and {Croll}, Bryce and {Shekhtman}, Louis M. and {Burrows}, Adam and {Deming}, Drake and {Greene}, Tom and {Hora}, Joseph L.},
        title = "{Thermal Phase Variations of WASP-12b: Defying Predictions}",
      journal = {\apj},
         year = 2012,
        month = mar,
       volume = {747},
       number = {1},
          eid = {82},
        pages = {82},
          doi = {10.1088/0004-637X/747/1/82},
archivePrefix = {arXiv},
       eprint = {1112.0574},
 primaryClass = {astro-ph.EP},
       adsurl = {https://ui.adsabs.harvard.edu/abs/2012ApJ...747...82C}
}

@ARTICLE{cowan13,
       author = {{Cowan}, Nicolas B. and {Fuentes}, Pablo A. and {Haggard}, Hal M.},
        title = "{Light curves of stars and exoplanets: estimating inclination, obliquity and albedo}",
      journal = {\mnras},
         year = 2013,
        month = sep,
       volume = {434},
       number = {3},
        pages = {2465-2479},
          doi = {10.1093/mnras/stt1191},
archivePrefix = {arXiv},
       eprint = {1304.6398},
 primaryClass = {astro-ph.EP},
       adsurl = {https://ui.adsabs.harvard.edu/abs/2013MNRAS.434.2465C}
}

@article{cullen24,
	adsurl = {https://ui.adsabs.harvard.edu/abs/2024MNRAS.531.1133C},
	author = {{Cullen}, Caitlyn J. and {Bayliss}, Daniel},
	doi = {10.1093/mnras/stae1197},
	journal = {\mnras},
	month = jun,
	number = {1},
	pages = {1133-1148},
	title = {{A search for non-transiting exoplanets with optical light phase curves from TESS Southern ecliptic sectors}},
	volume = {531},
	year = 2024}

@article{demory11,
	adsurl = {http://adsabs.harvard.edu/abs/2011ApJS..197...12D},
	archiveprefix = {arXiv},
	author = {{Demory}, B.-O. and {Seager}, S.},
	doi = {10.1088/0067-0049/197/1/12},
	eid = {12},
	eprint = {1110.6180},
	journal = {\apjs},
	month = nov,
	pages = {12},
	primaryclass = {astro-ph.EP},
	title = {{Lack of Inflated Radii for Kepler Giant Planet Candidates Receiving Modest Stellar Irradiation}},
	volume = 197,
	year = 2011}

@ARTICLE{demory13,
       author = {{Demory}, Brice-Olivier and {de Wit}, Julien and {Lewis}, Nikole and {Fortney}, Jonathan and {Zsom}, Andras and {Seager}, Sara and {Knutson}, Heather and {Heng}, Kevin and {Madhusudhan}, Nikku and {Gillon}, Michael and {Barclay}, Thomas and {Desert}, Jean-Michel and {Parmentier}, Vivien and {Cowan}, Nicolas B.},
        title = "{Inference of Inhomogeneous Clouds in an Exoplanet Atmosphere}",
      journal = {\apjl},
         year = 2013,
        month = oct,
       volume = {776},
       number = {2},
          eid = {L25},
        pages = {L25},
          doi = {10.1088/2041-8205/776/2/L25},
archivePrefix = {arXiv},
       eprint = {1309.7894},
 primaryClass = {astro-ph.EP},
       adsurl = {https://ui.adsabs.harvard.edu/abs/2013ApJ...776L..25D}
}

@article{esteves13,
	adsurl = {http://adsabs.harvard.edu/abs/2013ApJ...772...51E},
	archiveprefix = {arXiv},
	author = {{Esteves}, L.~J. and {De Mooij}, E.~J.~W. and {Jayawardhana}, R.},
	doi = {10.1088/0004-637X/772/1/51},
	eid = {51},
	eprint = {1305.3271},
	journal = {\apj},
	month = jul,
	pages = {51},
	primaryclass = {astro-ph.EP},
	title = {{Optical Phase Curves of Kepler Exoplanets}},
	volume = 772,
	year = 2013}

@article{esteves14,
	adsurl = {http://cdsads.u-strasbg.fr/abs/2015ApJ...804..150E},
	archiveprefix = {arXiv},
	author = {{Esteves}, L.~J. and {De Mooij}, E.~J.~W. and {Jayawardhana}, R.},
	doi = {10.1088/0004-637X/804/2/150},
	eid = {150},
	eprint = {1407.2245},
	journal = {\apj},
	month = may,
	pages = {150},
	primaryclass = {astro-ph.EP},
	title = {{Changing Phases of Alien Worlds: Probing Atmospheres of Kepler Planets with High-precision Photometry}},
	volume = 804,
	year = 2015}

@article{faigler11,
	adsurl = {http://adsabs.harvard.edu/abs/2011MNRAS.415.3921F},
	archiveprefix = {arXiv},
	author = {{Faigler}, S. and {Mazeh}, T.},
	doi = {10.1111/j.1365-2966.2011.19011.x},
	eprint = {1106.2713},
	journal = {\mnras},
	month = aug,
	pages = {3921-3928},
	primaryclass = {astro-ph.EP},
	title = {{Photometric detection of non-transiting short-period low-mass companions through the beaming, ellipsoidal and reflection effects in Kepler and CoRoT light curves}},
	volume = 415,
	year = 2011}

@article{faigler12,
	adsurl = {http://adsabs.harvard.edu/abs/2012ApJ...746..185F},
	archiveprefix = {arXiv},
	author = {{Faigler}, S. and {Mazeh}, T. and {Quinn}, S.~N. and {Latham}, D.~W. and {Tal-Or}, L.},
	doi = {10.1088/0004-637X/746/2/185},
	eid = {185},
	eprint = {1110.2133},
	journal = {\apj},
	month = feb,
	pages = {185},
	primaryclass = {astro-ph.EP},
	title = {{Seven New Binaries Discovered in the Kepler Light Curves through the BEER Method Confirmed by Radial-velocity Observations}},
	volume = 746,
	year = 2012}

@article{faigler13,
	adsurl = {http://adsabs.harvard.edu/abs/2013ApJ...771...26F},
	archiveprefix = {arXiv},
	author = {{Faigler}, S. and {Tal-Or}, L. and {Mazeh}, T. and {Latham}, D.~W. and {Buchhave}, L.~A.},
	doi = {10.1088/0004-637X/771/1/26},
	eid = {26},
	eprint = {1304.6841},
	journal = {\apj},
	month = jul,
	pages = {26},
	primaryclass = {astro-ph.EP},
	title = {{BEER Analysis of Kepler and CoRoT Light Curves. I. Discovery of Kepler-76b: A Hot Jupiter with Evidence for Superrotation}},
	volume = 771,
	year = 2013}

@article{faigler14,
	adsurl = {http://adsabs.harvard.edu/abs/2014arXiv1407.2361F},
	archiveprefix = {arXiv},
	author = {{Faigler}, S. and {Mazeh}, T.},
	eprint = {1407.2361},
	journal = {ArXiv e-prints},
	month = jul,
	primaryclass = {astro-ph.EP},
	title = {{BEER analysis of Kepler and CoRoT light curves: II. Evidence for emission phase shift due to superrotation in four Kepler hot Jupiters}},
	volume = {1407.2361},
	year = 2014}

@ARTICLE{gordon03,
       author = {{Walker}, Gordon and {Matthews}, Jaymie and {Kuschnig}, Rainer and {Johnson}, Ron and {Rucinski}, Slavek and {Pazder}, John and {Burley}, Gregory and {Walker}, Andrew and {Skaret}, Kristina and {Zee}, Robert and {Grocott}, Simon and {Carroll}, Kieran and {Sinclair}, Peter and {Sturgeon}, Don and {Harron}, John},
        title = "{The MOST Asteroseismology Mission: Ultraprecise Photometry from Space}",
      journal = {\pasp},
         year = 2003,
        month = sep,
       volume = {115},
       number = {811},
        pages = {1023-1035},
          doi = {10.1086/377358},
       adsurl = {https://ui.adsabs.harvard.edu/abs/2003PASP..115.1023W}
}

@article{greene01,
	adsurl = {https://ui.adsabs.harvard.edu/abs/2001ApJ...554.1290G},
	archiveprefix = {arXiv},
	author = {{Greene}, Jenny and {Bailyn}, Charles D. and {Orosz}, Jerome A.},
	doi = {10.1086/321411},
	eprint = {astro-ph/0101337},
	journal = {\apj},
	month = jun,
	number = {2},
	pages = {1290-1297},
	primaryclass = {astro-ph},
	title = {{Optical and Infrared Photometry of the Microquasar GRO J1655-40 in Quiescence}},
	volume = {554},
	year = 2001}

@article{gunther20,
	adsurl = {https://ui.adsabs.harvard.edu/abs/2020arXiv200314371G},
	archiveprefix = {arXiv},
	author = {{G{\"u}nther}, Maximilian N. and {Daylan}, Tansu},
	eid = {arXiv:2003.14371},
	eprint = {2003.14371},
	journal = {arXiv e-prints},
	month = mar,
	pages = {arXiv:2003.14371},
	primaryclass = {astro-ph.EP},
	title = {{Allesfitter: Flexible Star and Exoplanet Inference From Photometry and Radial Velocity}},
	year = 2020}

@ARTICLE{heng21,
       author = {{Heng}, Kevin and {Morris}, Brett M. and {Kitzmann}, Daniel},
        title = "{Closed-form ab initio solutions of geometric albedos and reflected light phase curves of exoplanets}",
      journal = {Nature Astronomy},
         year = 2021,
        month = oct,
       volume = {5},
        pages = {1001-1008},
          doi = {10.1038/s41550-021-01444-7},
archivePrefix = {arXiv},
       eprint = {2103.02673},
 primaryClass = {astro-ph.EP},
       adsurl = {https://ui.adsabs.harvard.edu/abs/2021NatAs...5.1001H}
}

@article{howell14,
	adsurl = {http://adsabs.harvard.edu/abs/2014PASP..126..398H},
	archiveprefix = {arXiv},
	author = {{Howell}, S.~B. and {Sobeck}, C. and {Haas}, M. and {Still}, M. and {Barclay}, T. and {Mullally}, F. and {Troeltzsch}, J. and {Aigrain}, S. and {Bryson}, S.~T. and {Caldwell}, D. and {Chaplin}, W.~J. and {Cochran}, W.~D. and {Huber}, D. and {Marcy}, G.~W. and {Miglio}, A. and {Najita}, J.~R. and {Smith}, M. and {Twicken}, J.~D. and {Fortney}, J.~J.},
	doi = {10.1086/676406},
	eprint = {1402.5163},
	journal = {\pasp},
	month = apr,
	pages = {398-408},
	primaryclass = {astro-ph.IM},
	title = {{The K2 Mission: Characterization and Early Results}},
	volume = 126,
	year = 2014}

@article{kane10,
	adsurl = {http://adsabs.harvard.edu/abs/2010ApJ...724..818K},
	archiveprefix = {arXiv},
	author = {{Kane}, S.~R. and {Gelino}, D.~M.},
	doi = {10.1088/0004-637X/724/1/818},
	eprint = {1009.4931},
	journal = {\apj},
	month = nov,
	pages = {818-826},
	primaryclass = {astro-ph.EP},
	title = {{Photometric Phase Variations of Long-period Eccentric Planets}},
	volume = 724,
	year = 2010}

@article{kane12,
	adsurl = {http://adsabs.harvard.edu/abs/2012MNRAS.424..779K},
	archiveprefix = {arXiv},
	author = {{Kane}, S.~R. and {Gelino}, D.~M.},
	doi = {10.1111/j.1365-2966.2012.21265.x},
	eprint = {1205.5812},
	journal = {\mnras},
	month = jul,
	pages = {779-788},
	primaryclass = {astro-ph.EP},
	title = {{Distinguishing between stellar and planetary companions with phase monitoring}},
	volume = 424,
	year = 2012}

@article{knutson07,
	adsurl = {http://adsabs.harvard.edu/abs/2007Natur.447..183K},
	archiveprefix = {arXiv},
	author = {{Knutson}, H.~A. and {Charbonneau}, D. and {Allen}, L.~E. and {Fortney}, J.~J. and {Agol}, E. and {Cowan}, N.~B. and {Showman}, A.~P. and {Cooper}, C.~S. and {Megeath}, S.~T.},
	doi = {10.1038/nature05782},
	eprint = {0705.0993},
	journal = {\nat},
	month = may,
	pages = {183-186},
	title = {{A map of the day-night contrast of the extrasolar planet HD 189733b}},
	volume = 447,
	year = 2007}

@article{knutson09,
	adsurl = {http://adsabs.harvard.edu/abs/2009ApJ...703..769K},
	archiveprefix = {arXiv},
	author = {{Knutson}, H.~A. and {Charbonneau}, D. and {Cowan}, N.~B. and {Fortney}, J.~J. and {Showman}, A.~P. and {Agol}, E. and {Henry}, G.~W.},
	doi = {10.1088/0004-637X/703/1/769},
	eprint = {0908.1977},
	journal = {\apj},
	month = sep,
	pages = {769-784},
	primaryclass = {astro-ph.EP},
	title = {{The 8 {$\mu$}m Phase Variation of the Hot Saturn HD 149026b}},
	volume = 703,
	year = 2009}

@book{lambert1760,
	address = {Zurich},
	publisher = {Heidegguer},
	adsurl = {http://adsabs.harvard.edu/abs/1759lpad.book.....L},
	author = {{Lambert}, J.~H.},
	booktitle = {Zurich : Heidegguer, 1759; VIII, 192 p.~: 6 tavv.~f.~t.~; in 8.; DCC.16.29},
	title = {{L perspective affranchie de l'embaras du Plan geometral}},
	year = 1759}

@article{lillo-box14,
	adsurl = {http://adsabs.harvard.edu/abs/2014A%26A...562A.109L},
	archiveprefix = {arXiv},
	author = {{Lillo-Box}, J. and {Barrado}, D. and {~Moya}, A. and {Montesinos}, B. and {Montalb{\'a}n}, J. and {Bayo}, A. and {Barbieri}, M. and {R{\'e}gulo}, C. and {Mancini}, L. and {Bouy}, H. and {Henning}, T.},
	doi = {10.1051/0004-6361/201322001},
	eid = {A109},
	eprint = {1312.3943},
	journal = {\aap},
	month = feb,
	pages = {A109},
	primaryclass = {astro-ph.EP},
	title = {{Kepler-91b: a planet at the end of its life. Planet and giant host star properties via light-curve variations}},
	volume = 562,
	year = 2014}

@article{lillo-box21,
	archiveprefix = {arXiv},
	author = {{Lillo-Box}, J. and {Millholland}, S. and {Laughlin}, G.},
	journal = {\aap},
	month = oct,
	primaryclass = {astro-ph.EP},
	title = {{Non-transiting planets from Kepler}},
	volume = {submitted},
	year = 2021}

@article{lillo-box21b,
	adsurl = {https://ui.adsabs.harvard.edu/abs/2021A&A...653A..40L},
	archiveprefix = {arXiv},
	author = {{Lillo-Box}, J. and {Ribas}, {\'A}. and {Montesinos}, B. and {Santos}, N.~C. and {Campante}, T. and {Cunha}, M. and {Barrado}, D. and {Villaver}, E. and {Sousa}, S. and {Bouy}, H. and {Aller}, A. and {Corsaro}, E. and {Li}, T. and {Ong}, J.~M.~J. and {Rebollido}, I. and {Audenaert}, J. and {Pereira}, F.},
	doi = {10.1051/0004-6361/202141158},
	eid = {A40},
	eprint = {2106.05011},
	journal = {\aap},
	month = sep,
	pages = {A40},
	primaryclass = {astro-ph.EP},
	title = {{Uncovering the ultimate planet impostor. An eclipsing brown dwarf in a hierarchical triple with two evolved stars}},
	volume = {653},
	year = 2021}

@article{loeb03,
	adsurl = {http://adsabs.harvard.edu/abs/2003ApJ...588L.117L},
	author = {{Loeb}, A. and {Gaudi}, B.~S.},
	doi = {10.1086/375551},
	eprint = {arXiv:astro-ph/0303212},
	journal = {\apjl},
	month = may,
	pages = {L117-L120},
	title = {{Periodic Flux Variability of Stars due to the Reflex Doppler Effect Induced by Planetary Companions}},
	volume = 588,
	year = 2003}

@article{madhusudhan12,
	adsurl = {http://adsabs.harvard.edu/abs/2012ApJ...747...25M},
	archiveprefix = {arXiv},
	author = {{Madhusudhan}, N. and {Burrows}, A.},
	doi = {10.1088/0004-637X/747/1/25},
	eid = {25},
	eprint = {1112.4476},
	journal = {\apj},
	month = mar,
	pages = {25},
	primaryclass = {astro-ph.EP},
	title = {{Analytic Models for Albedos, Phase Curves, and Polarization of Reflected Light from Exoplanets}},
	volume = 747,
	year = 2012}

@article{mazeh12,
	adsurl = {http://adsabs.harvard.edu/abs/2012A%26A...541A..56M},
	archiveprefix = {arXiv},
	author = {{Mazeh}, T. and {Nachmani}, G. and {Sokol}, G. and {Faigler}, S. and {Zucker}, S.},
	doi = {10.1051/0004-6361/201117908},
	eid = {A56},
	eprint = {1110.3512},
	journal = {\aap},
	month = may,
	pages = {A56},
	primaryclass = {astro-ph.EP},
	title = {{Kepler KOI-13.01 - Detection of beaming and ellipsoidal modulations pointing to a massive hot Jupiter}},
	volume = 541,
	year = 2012}

@article{millholland17,
	adsurl = {https://ui.adsabs.harvard.edu/abs/2017AJ....154...83M},
	archiveprefix = {arXiv},
	author = {{Millholland}, Sarah and {Laughlin}, Gregory},
	doi = {10.3847/1538-3881/aa7a0f},
	eid = {83},
	eprint = {1706.06602},
	journal = {\aj},
	month = {Sep},
	number = {3},
	pages = {83},
	primaryclass = {astro-ph.EP},
	title = {{Supervised Learning Detection of Sixty Non-transiting Hot Jupiter Candidates}},
	volume = {154},
	year = {2017}}

@article{mislis12,
	adsurl = {http://adsabs.harvard.edu/abs/2012A%26A...538A...4M},
	archiveprefix = {arXiv},
	author = {{Mislis}, D. and {Heller}, R. and {Schmitt}, J.~H.~M.~M. and {Hodgkin}, S.},
	doi = {10.1051/0004-6361/201116711},
	eid = {A4},
	eprint = {1112.2008},
	journal = {\aap},
	month = feb,
	pages = {A4},
	primaryclass = {astro-ph.EP},
	title = {{Estimating transiting exoplanet masses from precise optical photometry}},
	volume = 538,
	year = 2012}

@article{miszalski09,
	adsurl = {https://ui.adsabs.harvard.edu/abs/2009A&A...496..813M},
	archiveprefix = {arXiv},
	author = {{Miszalski}, B. and {Acker}, A. and {Moffat}, A.~F.~J. and {Parker}, Q.~A. and {Udalski}, A.},
	doi = {10.1051/0004-6361/200811380},
	eprint = {0901.4419},
	journal = {\aap},
	month = mar,
	number = {3},
	pages = {813-825},
	primaryclass = {astro-ph.SR},
	title = {{Binary planetary nebulae nuclei towards the Galactic bulge. I. Sample discovery, period distribution, and binary fraction}},
	volume = {496},
	year = 2009}

@article{morris85,
	adsurl = {http://adsabs.harvard.edu/abs/1985ApJ...295..143M},
	author = {{Morris}, S.~L.},
	doi = {10.1086/163359},
	journal = {\apj},
	month = aug,
	pages = {143-152},
	title = {{The ellipsoidal variable stars}},
	volume = 295,
	year = 1985}

@article{morris93,
	adsurl = {http://adsabs.harvard.edu/abs/1993ApJ...419..344M},
	author = {{Morris}, S.~L. and {Naftilan}, S.~A.},
	doi = {10.1086/173488},
	journal = {\apj},
	month = dec,
	pages = {344},
	title = {{The Equations of Ellipsoidal Star Variability Applied to HR 8427}},
	volume = 419,
	year = 1993}

@article{penoyre18,
	adsurl = {https://ui.adsabs.harvard.edu/abs/2019MNRAS.488.4181P},
	archiveprefix = {arXiv},
	author = {{Penoyre}, Zephyr and {Sandford}, Emily},
	doi = {10.1093/mnras/stz1941},
	eprint = {1803.07078},
	journal = {\mnras},
	month = sep,
	number = {3},
	pages = {4181-4194},
	primaryclass = {astro-ph.EP},
	title = {{Higher order harmonics in the light curves of eccentric planetary systems}},
	volume = {488},
	year = 2019}

@article{pfahl08,
	adsurl = {http://adsabs.harvard.edu/abs/2008ApJ...679..783P},
	archiveprefix = {arXiv},
	author = {{Pfahl}, E. and {Arras}, P. and {Paxton}, B.},
	doi = {10.1086/586878},
	eprint = {0704.1910},
	journal = {\apj},
	month = may,
	pages = {783-796},
	title = {{Ellipsoidal Oscillations Induced by Substellar Companions: A Prospect for the Kepler Mission}},
	volume = 679,
	year = 2008}

@article{phoebe,
	adsurl = {https://ui.adsabs.harvard.edu/abs/2016ApJS..227...29P},
	archiveprefix = {arXiv},
	author = {{Pr{\v{s}}a}, A. and {Conroy}, K.~E. and {Horvat}, M. and {Pablo}, H. and {Kochoska}, A. and {Bloemen}, S. and {Giammarco}, J. and {Hambleton}, K.~M. and {Degroote}, P.},
	doi = {10.3847/1538-4365/227/2/29},
	eid = {29},
	eprint = {1609.08135},
	journal = {\apjs},
	month = dec,
	number = {2},
	pages = {29},
	primaryclass = {astro-ph.SR},
	title = {{Physics Of Eclipsing Binaries. II. Toward the Increased Model Fidelity}},
	volume = {227},
	year = 2016}

@ARTICLE{placek14,
       author = {{Placek}, Ben and {Knuth}, Kevin H. and {Angerhausen}, Daniel},
        title = "{EXONEST: Bayesian Model Selection Applied to the Detection and Characterization of Exoplanets via Photometric Variations}",
      journal = {\apj},
         year = 2014,
        month = nov,
       volume = {795},
       number = {2},
          eid = {112},
        pages = {112},
          doi = {10.1088/0004-637X/795/2/112},
archivePrefix = {arXiv},
       eprint = {1310.6764},
 primaryClass = {astro-ph.EP},
       adsurl = {https://ui.adsabs.harvard.edu/abs/2014ApJ...795..112P}
}

@article{placek15,
	adsurl = {http://adsabs.harvard.edu/abs/2015ApJ...814..147P},
	archiveprefix = {arXiv},
	author = {{Placek}, B. and {Knuth}, K.~H. and {Angerhausen}, D. and {Jenkins}, J.~M.},
	doi = {10.1088/0004-637X/814/2/147},
	eid = {147},
	eprint = {1511.01068},
	journal = {\apj},
	month = dec,
	pages = {147},
	primaryclass = {astro-ph.EP},
	title = {{Characterization of Kepler-91b and the Investigation of a Potential Trojan Companion Using EXONEST}},
	volume = 814,
	year = 2015}

@article{quintana13,
	adsurl = {http://adsabs.harvard.edu/abs/2013ApJ...767..137Q},
	archiveprefix = {arXiv},
	author = {{Quintana}, E.~V. and {Rowe}, J.~F. and {Barclay}, T. and {Howell}, S.~B. and {Ciardi}, D.~R. and {Demory}, B.-O. and {Caldwell}, D.~A. and {Borucki}, W.~J. and {Christiansen}, J.~L. and {Jenkins}, J.~M. and {Klaus}, T.~C. and {Fulton}, B.~J. and {Morris}, R.~L. and {Sanderfer}, D.~T. and {Shporer}, A. and {Smith}, J.~C. and {Still}, M. and {Thompson}, S.~E.},
	doi = {10.1088/0004-637X/767/2/137},
	eid = {137},
	eprint = {1303.0858},
	journal = {\apj},
	month = apr,
	pages = {137},
	primaryclass = {astro-ph.EP},
	title = {{Confirmation of Hot Jupiter Kepler-41b via Phase Curve Analysis}},
	volume = 767,
	year = 2013}

@article{rauer14,
	adsurl = {http://ads.astro.puc.cl/abs/2014ExA....38..249R},
	archiveprefix = {arXiv},
	author = {{Rauer}, H. and {Catala}, C. and {Aerts}, C. and {Appourchaux}, T. and {Benz}, W. and {Brandeker}, A. and {Christensen-Dalsgaard}, J. and {Deleuil}, M. and {Gizon}, L. and {Goupil}, M.-J. and {G{\"u}del}, M. and {Janot-Pacheco}, E. and {Mas-Hesse}, M. and {Pagano}, I. and {Piotto}, G. and {Pollacco}, D. and {Santos}, {\.C}. and {Smith}, A. and {Su{\'a}rez}, J.-C. and {Szab{\'o}}, R. and {Udry}, S. and {Adibekyan}, V. and {Alibert}, Y. and {Almenara}, J.-M. and {Amaro-Seoane}, P. and {Eiff}, M.~A.-v. and {Asplund}, M. and {Antonello}, E. and {Barnes}, S. and {Baudin}, F. and {Belkacem}, K. and {Bergemann}, M. and {Bihain}, G. and {Birch}, A.~C. and {Bonfils}, X. and {Boisse}, I. and {Bonomo}, A.~S. and {Borsa}, F. and {Brand{\~a}o}, I.~M. and {Brocato}, E. and {Brun}, S. and {Burleigh}, M. and {Burston}, R. and {Cabrera}, J. and {Cassisi}, S. and {Chaplin}, W. and {Charpinet}, S. and {Chiappini}, C. and {Church}, R.~P. and {Csizmadia}, S. and {Cunha}, M. and {Damasso}, M. and {Davies}, M.~B. and {Deeg}, H.~J. and {D{\'{\i}}az}, R.~F. and {Dreizler}, S. and {Dreyer}, C. and {Eggenberger}, P. and {Ehrenreich}, D. and {Eigm{\"u}ller}, P. and {Erikson}, A. and {Farmer}, R. and {Feltzing}, S. and {de Oliveira Fialho}, F. and {Figueira}, P. and {Forveille}, T. and {Fridlund}, M. and {Garc{\'{\i}}a}, R.~A. and {Giommi}, P. and {Giuffrida}, G. and {Godolt}, M. and {Gomes da Silva}, J. and {Granzer}, T. and {Grenfell}, J.~L. and {Grotsch-Noels}, A. and {G{\"u}nther}, E. and {Haswell}, C.~A. and {Hatzes}, A.~P. and {H{\'e}brard}, G. and {Hekker}, S. and {Helled}, R. and {Heng}, K. and {Jenkins}, J.~M. and {Johansen}, A. and {Khodachenko}, M.~L. and {Kislyakova}, K.~G. and {Kley}, W. and {Kolb}, U. and {Krivova}, N. and {Kupka}, F. and {Lammer}, H. and {Lanza}, A.~F. and {Lebreton}, Y. and {Magrin}, D. and {Marcos-Arenal}, P. and {Marrese}, P.~M. and {Marques}, J.~P. and {Martins}, J. and {Mathis}, S. and {Mathur}, S. and {Messina}, S. and {Miglio}, A. and {Montalban}, J. and {Montalto}, M. and {Monteiro}, M.~J.~P.~F.~G. and {Moradi}, H. and {Moravveji}, E. and {Mordasini}, C. and {Morel}, T. and {Mortier}, A. and {Nascimbeni}, V. and {Nelson}, R.~P. and {Nielsen}, M.~B. and {Noack}, L. and {Norton}, A.~J. and {Ofir}, A. and {Oshagh}, M. and {Ouazzani}, R.-M. and {P{\'a}pics}, P. and {Parro}, V.~C. and {Petit}, P. and {Plez}, B. and {Poretti}, E. and {Quirrenbach}, A. and {Ragazzoni}, R. and {Raimondo}, G. and {Rainer}, M. and {Reese}, D.~R. and {Redmer}, R. and {Reffert}, S. and {Rojas-Ayala}, B. and {Roxburgh}, I.~W. and {Salmon}, S. and {Santerne}, A. and {Schneider}, J. and {Schou}, J. and {Schuh}, S. and {Schunker}, H. and {Silva-Valio}, A. and {Silvotti}, R. and {Skillen}, I. and {Snellen}, I. and {Sohl}, F. and {Sousa}, S.~G. and {Sozzetti}, A. and {Stello}, D. and {Strassmeier}, K.~G. and {{\v S}vanda}, M. and {Szab{\'o}}, G.~M. and {Tkachenko}, A. and {Valencia}, D. and {Van Grootel}, V. and {Vauclair}, S.~D. and {Ventura}, P. and {Wagner}, F.~W. and {Walton}, N.~A. and {Weingrill}, J. and {Werner}, S.~C. and {Wheatley}, P.~J. and {Zwintz}, K.},
	doi = {10.1007/s10686-014-9383-4},
	eprint = {1310.0696},
	journal = {Experimental Astronomy},
	month = nov,
	pages = {249-330},
	primaryclass = {astro-ph.EP},
	title = {{The PLATO 2.0 mission}},
	volume = 38,
	year = 2014}

@inproceedings{ricker14,
	adsurl = {http://adsabs.harvard.edu/abs/2014SPIE.9143E..20R},
	archiveprefix = {arXiv},
	author = {{Ricker}, G.~R. and {Winn}, J.~N. and {Vanderspek}, R. and {Latham}, D.~W. and {Bakos}, G.~{\'A}. and {Bean}, J.~L. and {Berta-Thompson}, Z.~K. and {Brown}, T.~M. and {Buchhave}, L. and {Butler}, N.~R. and {Butler}, R.~P. and {Chaplin}, W.~J. and {Charbonneau}, D. and {Christensen-Dalsgaard}, J. and {Clampin}, M. and {Deming}, D. and {Doty}, J. and {De Lee}, N. and {Dressing}, C. and {Dunham}, E.~W. and {Endl}, M. and {Fressin}, F. and {Ge}, J. and {Henning}, T. and {Holman}, M.~J. and {Howard}, A.~W. and {Ida}, S. and {Jenkins}, J. and {Jernigan}, G. and {Johnson}, J.~A. and {Kaltenegger}, L. and {Kawai}, N. and {Kjeldsen}, H. and {Laughlin}, G. and {Levine}, A.~M. and {Lin}, D. and {Lissauer}, J.~J. and {MacQueen}, P. and {Marcy}, G. and {McCullough}, P.~R. and {Morton}, T.~D. and {Narita}, N. and {Paegert}, M. and {Palle}, E. and {Pepe}, F. and {Pepper}, J. and {Quirrenbach}, A. and {Rinehart}, S.~A. and {Sasselov}, D. and {Sato}, B. and {Seager}, S. and {Sozzetti}, A. and {Stassun}, K.~G. and {Sullivan}, P. and {Szentgyorgyi}, A. and {Torres}, G. and {Udry}, S. and {Villasenor}, J.},
	booktitle = {Society of Photo-Optical Instrumentation Engineers (SPIE) Conference Series},
	doi = {10.1117/12.2063489},
	eid = {914320},
	eprint = {1406.0151},
	month = aug,
	pages = {20},
	primaryclass = {astro-ph.EP},
	series = {Society of Photo-Optical Instrumentation Engineers (SPIE) Conference Series},
	title = {{Transiting Exoplanet Survey Satellite (TESS)}},
	volume = 9143,
	year = 2014}

@article{russell1916,
	adsurl = {http://adsabs.harvard.edu/abs/1916ApJ....43..173R},
	author = {{Russell}, H.~N.},
	doi = {10.1086/142244},
	journal = {\apj},
	month = apr,
	pages = {173},
	title = {{On the Albedo of the Planets and Their Satellites}},
	volume = 43,
	year = 1916}

@ARTICLE{shakura87,
       author = {{Shakura}, N.~I. and {Postnov}, K.~A.},
        title = "{Doppler-effect modulation of the observed radiation flux from ultracompact binary stars.}",
      journal = {\aap},
         year = 1987,
        month = sep,
       volume = {183},
        pages = {L21-L22},
       adsurl = {https://ui.adsabs.harvard.edu/abs/1987A&A...183L..21S}
}

@article{showman02,
	adsurl = {http://adsabs.harvard.edu/abs/2002A%26A...385..166S},
	author = {{Showman}, A.~P. and {Guillot}, T.},
	doi = {10.1051/0004-6361:20020101},
	eprint = {astro-ph/0202236},
	journal = {\aap},
	month = apr,
	pages = {166-180},
	title = {{Atmospheric circulation and tides of ``51 Pegasus b-like'' planets}},
	volume = 385,
	year = 2002}

@ARTICLE{shporer14,
       author = {{Shporer}, Avi and {O'Rourke}, Joseph G. and {Knutson}, Heather A. and {Szab{\'o}}, Gyula M. and {Zhao}, Ming and {Burrows}, Adam and {Fortney}, Jonathan and {Agol}, Eric and {Cowan}, Nicolas B. and {Desert}, Jean-Michel and {Howard}, Andrew W. and {Isaacson}, Howard and {Lewis}, Nikole K. and {Showman}, Adam P. and {Todorov}, Kamen O.},
        title = "{Atmospheric Characterization of the Hot Jupiter Kepler-13Ab}",
      journal = {\apj},
         year = 2014,
        month = jun,
       volume = {788},
       number = {1},
          eid = {92},
        pages = {92},
          doi = {10.1088/0004-637X/788/1/92},
archivePrefix = {arXiv},
       eprint = {1403.6831},
 primaryClass = {astro-ph.EP},
       adsurl = {https://ui.adsabs.harvard.edu/abs/2014ApJ...788...92S}
}

@ARTICLE{shporer15,
       author = {{Shporer}, Avi and {Hu}, Renyu},
        title = "{Studying Atmosphere-dominated Hot Jupiter Kepler Phase Curves: Evidence that Inhomogeneous Atmospheric Reflection Is Common}",
      journal = {\aj},
         year = 2015,
        month = oct,
       volume = {150},
       number = {4},
          eid = {112},
        pages = {112},
          doi = {10.1088/0004-6256/150/4/112},
archivePrefix = {arXiv},
       eprint = {1504.00498},
 primaryClass = {astro-ph.SR},
       adsurl = {https://ui.adsabs.harvard.edu/abs/2015AJ....150..112S}
}

@article{southworth08,
	adsurl = {http://adsabs.harvard.edu/abs/2008MNRAS.386.1644S},
	archiveprefix = {arXiv},
	author = {{Southworth}, J.},
	doi = {10.1111/j.1365-2966.2008.13145.x},
	eprint = {0802.3764},
	journal = {\mnras},
	month = may,
	pages = {1644-1666},
	title = {{Homogeneous studies of transiting extrasolar planets - I. Light-curve analyses}},
	volume = 386,
	year = 2008}

@ARTICLE{spiegel11,
       author = {{Spiegel}, David S. and {Burrows}, Adam and {Milsom}, John A.},
        title = "{The Deuterium-burning Mass Limit for Brown Dwarfs and Giant Planets}",
      journal = {\apj},
         year = 2011,
        month = jan,
       volume = {727},
       number = {1},
          eid = {57},
        pages = {57},
          doi = {10.1088/0004-637X/727/1/57},
archivePrefix = {arXiv},
       eprint = {1008.5150},
 primaryClass = {astro-ph.EP},
       adsurl = {https://ui.adsabs.harvard.edu/abs/2011ApJ...727...57S}
}

@article{sudarsky05,
	adsurl = {http://adsabs.harvard.edu/abs/2005ApJ...627..520S},
	author = {{Sudarsky}, D. and {Burrows}, A. and {Hubeny}, I. and {Li}, A.},
	doi = {10.1086/430206},
	eprint = {astro-ph/0501109},
	journal = {\apj},
	month = jul,
	pages = {520-533},
	title = {{Phase Functions and Light Curves of Wide-Separation Extrasolar Giant Planets}},
	volume = 627,
	year = 2005}

@article{welsh10,
	adsurl = {http://adsabs.harvard.edu/abs/2010ApJ...713L.145W},
	archiveprefix = {arXiv},
	author = {{Welsh}, W.~F. and {Orosz}, J.~A. and {Seager}, S. and {Fortney}, J.~J. and {Jenkins}, J. and {Rowe}, J.~F. and {Koch}, D. and {Borucki}, W.~J.},
	doi = {10.1088/2041-8205/713/2/L145},
	eprint = {1001.0413},
	journal = {\apjl},
	month = apr,
	pages = {L145-L149},
	primaryclass = {astro-ph.SR},
	title = {{The Discovery of Ellipsoidal Variations in the Kepler Light Curve of HAT-P-7}},
	volume = 713,
	year = 2010}

@ARTICLE{wilson90,
       author = {{Wilson}, R.~E.},
        title = "{Accuracy and Efficiency in the Binary Star Reflection Effect}",
      journal = {\apj},
         year = 1990,
        month = jun,
       volume = {356},
        pages = {613},
          doi = {10.1086/168867},
       adsurl = {https://ui.adsabs.harvard.edu/abs/1990ApJ...356..613W}
}
\end{document}